\documentclass[%
 aip,
 amsmath,amssymb,
preprint,%
]{revtex4-1}

\usepackage{graphicx}
\usepackage{dcolumn}
\usepackage{bm}

\usepackage[utf8]{inputenc}
\usepackage[T1]{fontenc}
\usepackage{longtable} 
\usepackage{multirow} 
\usepackage{amsmath} 
\usepackage[T1]{fontenc} 
\usepackage{tabularx} 
\usepackage{changepage} 
\usepackage{natbib} 
\usepackage{float} 
\usepackage[running]{lineno} 
\usepackage{xcolor} 
\usepackage{relsize} 
\usepackage{url} 
\usepackage{appendix} 

\begin{document}
\preprint{}

\title{Level-crossings reveal organized coherent structures in a turbulent time series}

\author{Subharthi Chowdhuri}
\email{subharc@uci.edu}
\affiliation{Department of Civil and Environmental Engineering, University of California, Irvine, CA 92697, USA}

\author{Tirtha Banerjee}
\affiliation{Department of Civil and Environmental Engineering, University of California, Irvine, CA 92697, USA}

\date{\today}

\begin{abstract}
In turbulent flows, energy production is associated with highly organized structures, known as coherent structures. Since these structures are three-dimensional, their detection remains challenging in the most common situation, when single-point temporal measurements are considered. While previous research on coherent structure detection from time series employs a thresholding approach, the thresholds are ad-hoc and vary significantly from one study to another. To eliminate this subjective bias, we introduce the level-crossing method and show how specific features of a turbulent time series associated with coherent structures can be objectively identified, without assigning a prior any arbitrary threshold. By using two wall-bounded turbulence time series datasets, we successfully extract through level-crossing analysis the impacts of coherent structures on turbulent dynamics, and therefore, open an alternative avenue in experimental turbulence research. By utilizing this framework further we identify a new metric, characterized by a statistical asymmetry between peaks and troughs of a turbulent signal, to quantify inner-outer interaction in wall turbulence. Moreover, a connection is established between extreme value statistics and level-crossing analysis, thereby allowing additional possibilities to study extreme events in other dynamical systems.
\end{abstract}

\maketitle

\section{Introduction}
\label{Intro}
Coherent structures in turbulent flows, ranging from astrophysical to engineered, to atmospheric systems, are best described by their phenomenology, such as: (a) their characteristic scales are comparable to the integral scales \citep{kaftori1995particle}; (b) they induce non-Gaussian fluctuations in the turbulent variables \citep{majda1999simplified}; and (c) they have large contributions to turbulent fluxes and kinetic energy \citep{jimenez2018coherent}. Geometrically, these structures are three-dimensional and can take various shapes based on the types of turbulent flow. Examples include, granular patterns in astrophysical flows \citep{chian2014detection}; hairpin structures in neutral wall-bounded flows \citep{adrian2007hairpin}; and counter-rotating roll vortices in atmospheric turbulence \citep{young2002rolls}. Notwithstanding their significance in drag reduction \citep{marusic2021energy}, consideration of coherent structures are also important for weather and climate models since disregarding those can cause significant uncertainties in turbulence parameterization \citep{salesky2020coherent}.

Despite these structures can be visually recognized from three-dimensional numerical simulations, smoke visualization experiments, particle velocimetry measurements, or satellite images, it remains challenging to detect them from the most common form of turbulent experiments where the variables are measured at a single point in time. Previous studies on coherent structure detection from turbulent time series employ a thresholding approach, where the thresholds are set either in the temporal or spectral domain. 

Regarding spectral domain, \citet{perry1982mechanism} demonstrated how the spectra of streamwise velocity fluctuations in time encoded the information about hairpin eddy structures by displaying a $-1$ spectral power law. Therefore, by choosing an appropriate cut-off wavelength ($\lambda$), the effect of hairpin eddy structures on the velocity statistics could be inferred from a single point time series \citep{marusic2019attached}. Apart from hairpin eddies, the contributions from very large scale motions (VLSMs) on velocity statistics are detected by setting $\lambda$ comparable to the boundary layer depth \citep{balakumar2007large}. By contrast, in the temporal domain, thresholds are applied directly on the time series and historically were chosen in a manner so that the frequency of the detected structures matched with the smoke visualization experiments \citep{antonia1981conditional}.

However, the thresholding procedure in temporal domain suffers from subjectivity as the threshold values differ significantly from one study to another \citep{subramanian1982comparison}. Additionally, the rationale behind their choices also varies, since some studies consider the thresholds where the probability density functions (PDFs) of the time series differ from a Gaussian, whereas others choose them from a quadrant perspective \citep{alfonsi2006coherent}. Conversely, the thresholds in the spectral domain often require information about certain parameters (such as boundary layer height) whose measurements are rarely available \citep{liu2021large}. To eliminate these difficulties, we introduce a level-crossing technique through which the coherent structures can be detected from time series without assigning a prior any tunable thresholds or external parameters.

Although a handful of previous research, such as the ones by \citet{tardu2015level} and \citet{poggi2010evaluation}, have used level-crossing analysis to study the Reynolds stress production and dissipation of turbulence kinetic energy in wall turbulence, we demonstrate that this approach could be generalized further to detect coherent structures. In a level-crossing method \citep{blake1973level}, one seeks a statistical description of time scales $t_p\vert_{\alpha}$ up to which a stochastic variable $f(t)$ remains larger or smaller than $\overline{f}\pm(\alpha \times \sigma_f)$, where $t$ is time, $\overline{f}$ is the temporal mean, $\sigma_f$ is the standard deviation, and $\alpha$ is a given threshold. A brief review of level-crossing approach is provided by \citet{friedrich2011approaching}, which, in other words, is a generalization of the zero-crossing or persistence analysis where $\alpha$ level is set at zero \citep{majumdar1999persistence,tardu2015level}. For many different turbulent flows, the PDFs ($P(t_p\vert_{\alpha=0})$) of $t_p\vert_{\alpha=0}$ are power-laws with an exponential cut-off \citep{perlekar2011persistence,chowdhuri2020persistence,heisel2022self}. On the one hand, the exponential cutoff represents a Poisson distribution, associated with $t_p$ values larger than the integral scales $\gamma$ \citep{cava2012role}. On the other hand, \citet{blake1973level} shows $P(t_p\vert_{\alpha})$ becomes a Poisson distribution when $\alpha$ values are substantially large. 

Several points are now considered. First, the Poisson distribution is associated with a stochastic process for which the autocorrelation function stays zero at all time scales \citep{majumdar1999persistence}. Second, in a turbulent time series, the measurements become weakly correlated with each other at scales larger than $\gamma$, since the autocorrelation functions drop to zero \citep{cava2012role}. Third, the characteristic scales of coherent structures are comparable to $\gamma$ \citep{kaftori1995particle}. Fourth, in a randomly-shuffled (RS) signal, the autocorrelation functions cease to exist \citep{lancaster2018surrogate}. By combining all these aspects, we hypothesize the thresholds to detect coherent structures could be objectively determined as that particular $\alpha$ for which $P(t_p\vert_{\alpha})$ of the original signal matches with its RS counterpart. Intrigued by this possibility, we ask: (1) By changing $\alpha$ what turbulent flow features are revealed? (2) Do the detected coherent structures from the critical $\alpha$ value obey the flow physics? (3) Can we identify the organizational aspects of coherent structures through the level-crossing approach? 

We focus our attention on wall-bounded turbulence, since in such flows the properties of coherent structures are well-established \citep{jimenez2018coherent}. We use two hot-wire temporal datasets, collected from a zero-pressure gradient turbulent boundary layer generated in the Melbourne wind tunnel \citep{MARUSIC2020dataset}. During our presentation, we arrange the article in three different sections. In Section \ref{Data}, we provide brief descriptions of the experimental datasets and methodology, in Section \ref{results} we introduce the results and discuss them, and lastly in Section \ref{conclusion} we summarize the key takeaways and provide the scope for further research.

\section{Dataset and methodology}
\label{Data}
Corresponding to both hot-wire datasets, the friction Reynolds numbers ($Re$) are of the order of $10^4$ as illustrated in \citet{baars2015wavelet}. The wall-normal heights are normalized by friction velocity ($u_{*}$) and kinematic viscosity ($\nu$) and denoted as $y^{+}$, where $^{+}$ refers to wall-scaling. We restrict $y^{+} \leq 10^4$, up to which the flow is fully turbulent \citep{iacobello2023coherent}. Out of the two datasets, one of them were sampled at a frequency ($f_s$) of 20 kHz (T1 dataset), while the other at 44 kHz (T2 dataset). Note that for both datasets, one probe is fixed (reference probe) while the others traverse across heights and are synchronized with the reference probe \citep{baars2015wavelet}. Moreover, for the T1 dataset, the time series of streamwise velocity were collected over three acquisition cycles, with each of $120$-s duration. Therefore, the results presented for the T1 dataset are ensemble-averaged over these three measurement cycles. However, for the T2 dataset, only a single cycle of $360$-s duration was used. Regarding our purposes, we consider the streamwise velocity fluctuations ($u^{\prime}$) after subtracting the temporal mean ($\overline{u}$). Subsequently, on these $u^{\prime}$ time-series we apply level-crossing and event-synchronization analysis, whose rationale are discussed below.  

\subsection{Level-crossing analysis}
\label{lc}
\begin{figure*}[h]
\centerline{\includegraphics[width=\textwidth]{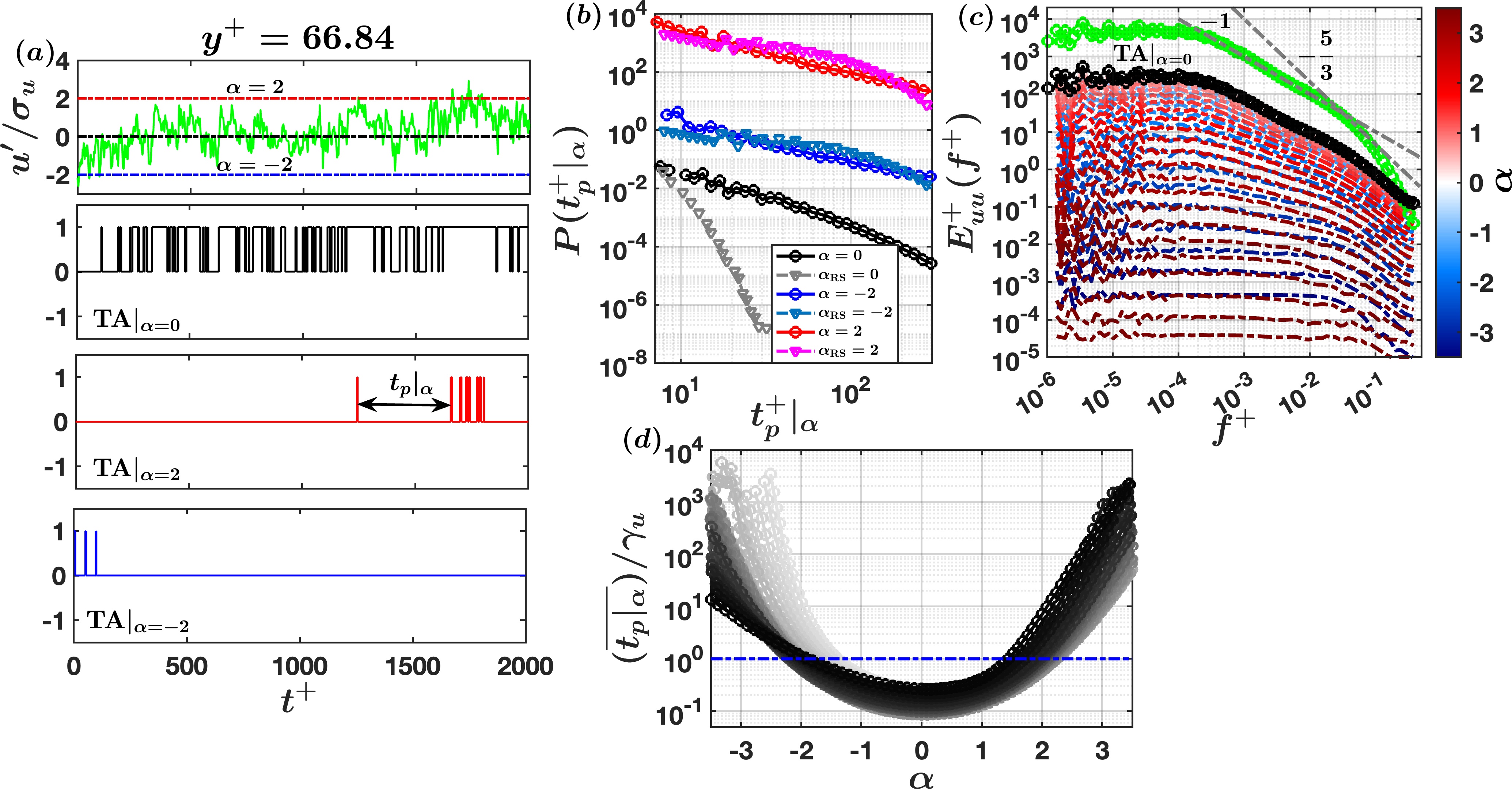}}
\caption{(a) A segment of $u^{\prime}$ time series and it’s telegraphic approximations (TA) at different threshold levels ($\alpha$) are shown from $y^{+}=$ 66.84 of T1 dataset. The level-crossing time scales at different $\alpha$ values are denoted as $t_{p}\vert_{\alpha}$. (b) The PDFs of $t^{+}_{p}\vert_{\alpha}$ are shown for $\alpha=0,2,-2$, corresponding to the original and randomly-shuffled (RS) signals. (c) The energy spectrum of the time series at $y^{+}=$ 66.84 is compared with the TA series at different $\alpha$ levels. The power-laws $-1$ and $-5/3$ are shown in dash-dotted gray lines. (d) Normalized mean time scales ($\overline{t_{p}\vert_{\alpha}}/\gamma_{u}$) are plotted against $\alpha$ values, by considering all the heights from the T1 dataset. The gray shaded colors indicate different heights as per the color-bar denoting $\log_{10}(y^{+})$ in Fig. \ref{fig:3}. The horizontal blue dash-dotted line indicates $\overline{t_{p}\vert_{\alpha}}=\gamma_{u}$.}
\label{fig:1}
\end{figure*}
To demonstrate the philosophy behind level-crossing analysis, we use a segment of a $u^{\prime}$ time series (normalized by its standard deviation $\sigma_u$) from the T1 dataset at height $y^{+}=66.84$ (Fig.\ref{fig:1}a). Corresponding to this time series, one can generate its telegraphic approximations (TA) by denoting the values above a threshold to be 1 and 0 otherwise \citep{sreenivasan2006clustering}. In the bottom panels of Fig.\ref{fig:1}a, we show three TA sequences at threshold levels $\alpha=0,2,-2$. One can clearly see as the threshold levels are increased, the timescales $t_p\vert_{\alpha}$ of the TA patterns become substantially large. In fact, if $t_p$ values become comparable to the integral scales of $u^{\prime}$ ($\gamma_u$), one would expect the TA patterns associated with those $\alpha$ levels to resemble a random configuration. 

This indeed appears to be the case when one compares the PDFs of $t_p\vert_{\alpha}$ for the three $\alpha$ values between the original and randomly-shuffled (RS) signals (Fig.\ref{fig:1}b). As opposed to $\alpha=0$, for $\alpha=2,-2$, $P(t_p\vert_{\alpha})$ of $u^{\prime}$ signal has an excellent agreement with its RS counterpart. This can be further confirmed through the q-q plots, where the $t_p\vert_{\alpha}$ values between the original and RS signals follow a straight line with 45$^{\circ}$ slope for $\alpha=-2,2$, thereby indicating they are both sampled from similar distributions (not shown here). 

It is also interesting to investigate how the energy spectrum of the TA patterns change as $\alpha$ is varied systematically. \citet{sreenivasan2006clustering} showed that the energy spectrum of TA patterns corresponding to $\alpha=0$ level, preserve the information about the spectral power laws albeit with some change depending on the Reynolds number of the flow. In Fig.\ref{fig:1}c we show the energy spectra of the TA patterns with different $\alpha$ values and compare the same with the original $u^{\prime}$ signal at $y^{+}=66.84$ (green line with circular markers). One can see at $\alpha=0$ level (black line with circles), the energy spectrum shows a $-1$ spectral scaling at smaller frequencies similar to original signal, but at larger frequencies the $-5/3$ scaling law appears to be a little different. However, at large enough $\alpha$ values (indicated by deep red or blue colors), the scaling laws disappear from the TA energy spectra and they nearly attain a flat shape as expected for a random signal. 

More importantly, for all the available heights from the T1 dataset, if one plots the mean time scales ($\overline{t_{p}\vert_{\alpha}}$) against the $\alpha$ values, then $\overline{t_{p}\vert_{\alpha}}$ exceeds $\gamma_u$ considerably as $\alpha$ increases (Fig.\ref{fig:1}d). Therefore, one can conclude that by increasing $\alpha$ a critical value is reached, using which one could study certain flow features whose characteristic scales are comparable to $\gamma_u$.  

\subsection{Event-Synchronization analysis}
\label{ES}
\begin{figure*}[h]
\centerline{\includegraphics[width=\textwidth]{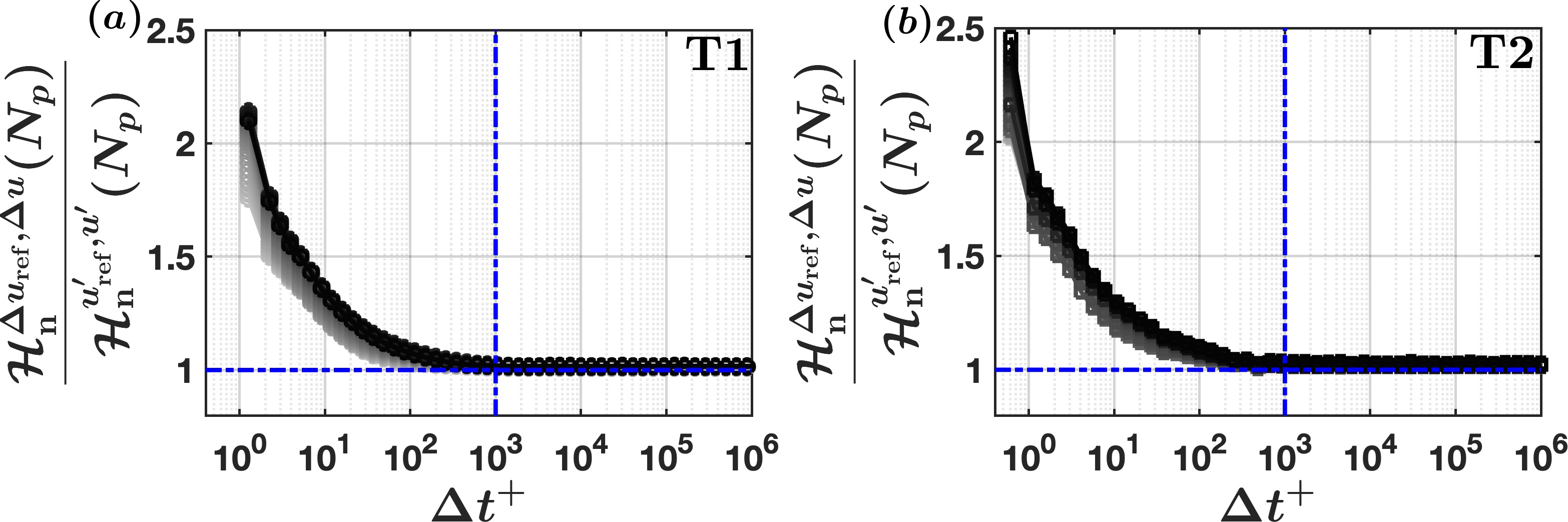}}
\caption{Shannon entropies of the synchronized event lengths are plotted against different time lags ($\Delta t^{+}$) for the (a) T1 and (b) T2 datasets, where the values are normalized by the entropies of the synchronized event lengths computed for the full signal. }
\label{fig:2}
\end{figure*}
By conducting an event-synchronization analysis, one seeks to describe how the positive and negative patterns (with respect to $\alpha=0$ level) in a turbulent signal are coupled with each other across different wall-normal heights. This information is important to establish how the non-local influences impact the organization of turbulent events in wall-bounded flows. 

Before explaining this analysis it is prudent to mention briefly about the probe arrangements for the T1 and T2 datasets. For both datasets, one probe is fixed at a location (either in the inner or outer layer), whereas the other probes travel across $y^+$ while at all times being synchronized with the reference probe. Specifically, for the T1 dataset, the reference probe is located at $y^{+}=4.3$, and for T2 dataset it is at $y^{+}=474$ where the outer peaks appear in the turbulence spectra \citep{baars2015wavelet}.

For event synchronization analysis, we consider a joint distribution between the positive and negative patterns corresponding to the velocity signals from the reference probe ($u^{\prime}_{\rm ref}$) and from a traveling probe situated at any particular height ($u^{\prime}$). This joint distribution is studied in terms of a binary sequence whose values are 1 when $u^{\prime}_{\rm ref}$ and $u^{\prime}$ are simultaneously positive or negative. On the other hand, when the signs mismatch, the sequence attains zero. We refer to this as an overlap binary sequence and compute its time scales ($t_p$) based on the duration where it stays at 1 or 0. To quantify the synchronization, Shannon entropies of the overlap event lengths ($N_p$) are considered and compared with a RS sequence (by taking a ratio), which is supposedly devoid of any coupling effect. Mathematically, Shannon entropy of the overlap event lengths is bounded within, $0 \leq H^{x_{\rm ref},x}_{\rm n}(N_p) \leq 1$, where $x_{\rm ref}$ is the reference signal, $x$ is the signal from the travelling probe, and 1 (0) indicates no (complete) synchronization between the two signals.

To incorporate the effect of turbulent scales, the aforementioned procedure is operated on $\Delta u_{\rm ref}$ and $\Delta u$ signals, where $\Delta u$ denotes velocity differences at a time lag $\Delta t$. The time lags are normalized with wall-scaling and denoted as $\Delta t^{+}$. The synchronized entropy values at any $\Delta t^{+}$ are scaled with the entropy values for the full signal ($H^{u^{\prime}_{\rm ref},u^{\prime}}_{\rm n}(N_p)$). From Fig. \ref{fig:2} one can see that irrespective of the datasets or $y^{+}$ values, the scaled entropies decrease with increasing scales and approach unity at $\Delta t^{+}=10^3$, which physically represents the time scales of the outer- layer structures \citep{baars2015wavelet}. This implies at scales comparable to outer-layer structures the synchronized entropies of the velocity differences become equal to the full signal values. Therefore, one could infer that the positive and negative patterns in the $u^{\prime}$ signals at any $y^{+}$ value carry the signatures of the structures residing in the outer-layer. Further implications of this phenomenon are discussed in Section \ref{results}.

\section{Results and discussion}
\label{results}
\subsection{Level crossings and extreme values}
\label{Res1}
\begin{figure*}
\centerline{\includegraphics[width=\textwidth]{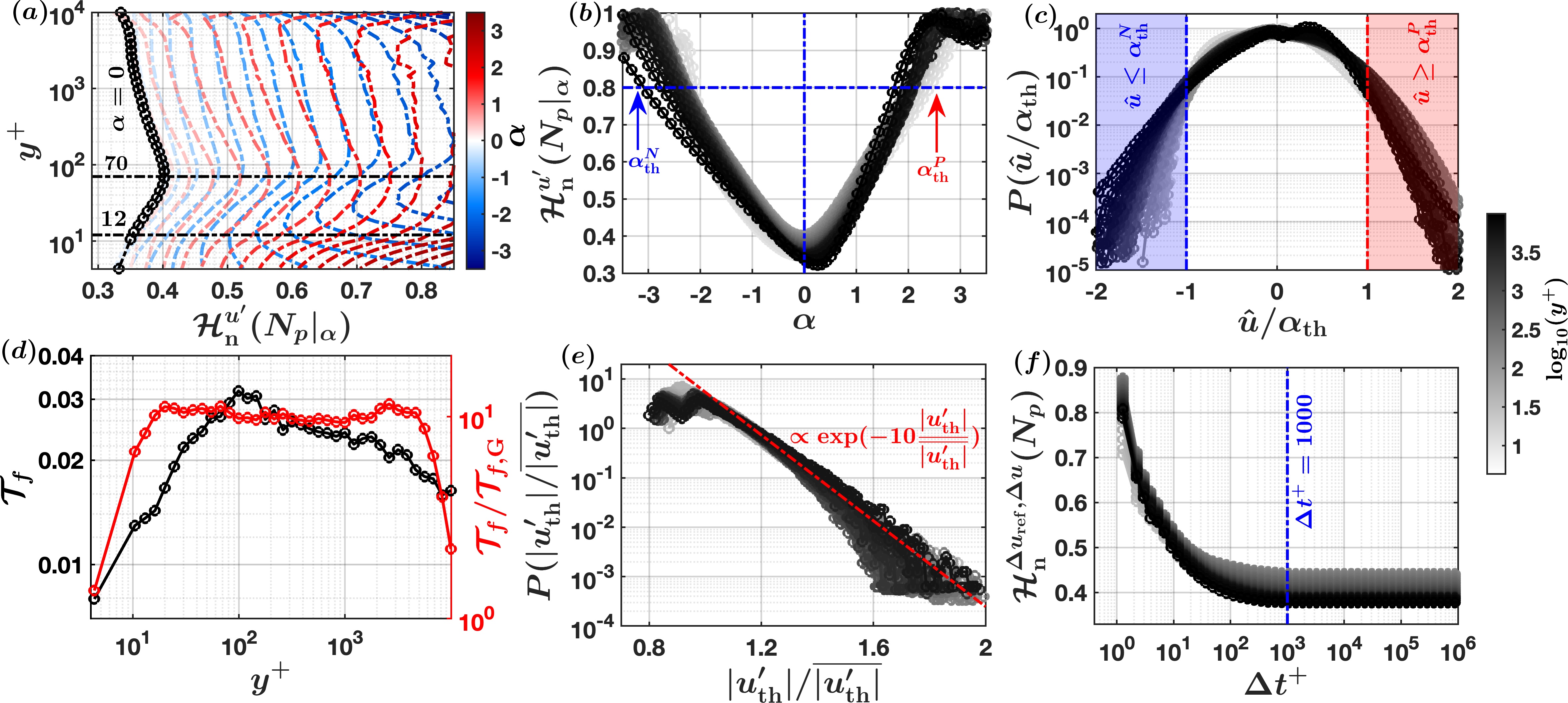}}
\caption{From the T1 dataset (a) vertical profiles of scaled Shannon entropies, corresponding to the event lengths at different $\alpha$ levels ($N_{p}\vert_{\alpha}$) are shown; (b) $\mathcal{H}^{u^{\prime}}_{\rm n}(N_{p}\vert_{\alpha})$ values are plotted for different $y^{+}$ values (see the color bar), and $\alpha^{N}_{\rm th}$ and $\alpha^{P}_{\rm th}$ are identified where $\mathcal{H}{u^{\prime}}_{\rm n}(N_{p}\vert_{\alpha})=0.8$; (c) PDFs of $\hat{u}/\alpha_{\rm th}$ are shown, where $\hat{u}=u^{\prime}/\sigma_{u}$; (d) time fractions ($\mathcal{T}_{f}$) occupied by the samples exceeding $\alpha_{th}$ and their comparison with a Gaussian distribution ($\mathcal{T}_{f,\rm G}$) is shown; (e) PDFs of $|u^{\prime}_{\rm th}|/\overline{|u^{\prime}_{\rm th}|}$ are shown, where $u^{\prime}_{\rm th}$ are the samples exceeding $\alpha_{\rm th}$; and (f) Shannon entropies of the synchronized event lengths are plotted against different time lags ($\Delta t^{+}$) and $\Delta u$ denotes velocity increments.}  
\label{fig:3}
\end{figure*}

To address the research questions, we begin with the probability distributions of event lengths ($P(N_p \vert_{\alpha})$) as $\alpha$ is varied. We consider $N_p \vert_{\alpha}$ since it is a discrete variable and represented through probability mass functions whose computation is insensitive to binning. Note that $N_p \vert_{\alpha}$ and $t_p \vert_{\alpha}$ are interchangeable through $t_p\vert_{\alpha}=N_p\vert_{\alpha}/f_s$. To characterize $P(N_p \vert_{\alpha})$, we consider its Shannon entropy compared with a RS sequence of $u^{\prime}$. The entropy is denoted as $H^{u^{\prime}}_{\rm n}(N_{p}\vert_{\alpha})$, whose mathematical expression is provided in Eq. (\ref{SN}) of Appendix \ref{app_A}. Furthermore, $H^{u^{\prime}}_{\rm n}(N_{p}\vert_{\alpha})$ is bounded between $0 \leq H^{u^{\prime}}_{\rm n}(N_{p}\vert_{\alpha}) \leq 1$ with 1 indicating a random configuration. From Fig. \ref{fig:3}a, one observes that the effect of changing $\alpha$ either from positive or negative side on $H^{u^{\prime}}_{\rm n}(N_{p}\vert_{\alpha})$ is asymmetric. The vertical profiles of $H^{u^{\prime}}_{\rm n}(N_{p}\vert_{\alpha \geq 0})$ show an inflection point around $y^{+} \approx 70$, while when $\alpha$ is approached from the negative side an another inflection point appears at $y^{+} \approx 12$. The position $y^{+}=70$ indicates the location where the outer layer begins \citep{bae2021life}, whereas at $y^{+}=12$ the inner-layer structures are active \citep{baars2015wavelet}. As shown later in Section \ref{Res3}, this asymmetrical progression is related to inner-outer interaction in wall turbulence.

However, these inflection points disappear with increasing $\alpha$. In fact, $H^{u^{\prime}}_{\rm n}(N_{p}\vert_{\alpha})$ tend towards unity at large $\alpha$ values (Fig. \ref{fig:3}b). This apparent randomness in $N_p$ is associated with the fact that with increasing $\alpha$, $t_p\vert_{\alpha}$ become statistically comparable to $\gamma_u$ (Fig. \ref{fig:1}d). For accuracy purposes, we consider those $\alpha$ values as the critical ones where $H^{u^{\prime}}_{\rm n}(N_{p}\vert_{\alpha})$ crosses 0.8 (see Appendix \ref{app_A}). These values are denoted as $\alpha^{P}_{\rm th}$ and $\alpha^{N}_{\rm th}$ (Fig. \ref{fig:3}b), respectively, and any difference between them is correlated to the skewness of $u^{\prime}$ ($\mathcal{S}(u')$, Fig. \ref{fig:5}b). 

Upon considering $P(u^{\prime})$, one can see the samples that exceed these critical values reside in the PDF tails (Fig. \ref{fig:3}c). For visualization purposes, before plotting $P(u^{\prime})$, we scale the positive and negative $u^{\prime}$ values with $\alpha^P_{\rm th}\sigma_{u}$ and $\alpha^N_{\rm th}\sigma_{u}$, respectively. Under this scaling, the values beyond $\pm 1$ in Fig. \ref{fig:3}c, indicate those critical $u^{\prime}$ samples exceeding either $\alpha^P_{\rm th}\sigma_{u}$ (red-shaded regions) or $\alpha^N_{\rm th}\sigma_{u}$ (blue-shaded regions). Specifically, from Fig. \ref{fig:3}d, the time fractions ($\mathcal{T}_f$) associated with these critical samples ($u^{\prime}_{\rm th}$) are nearly 1-3\% of the total sample length, and their values differ significantly from the ones obtained through a Gaussian distribution of $u^{\prime}$ ($\mathcal{T}_{f,\rm G}$). Accordingly, $P(|u^{\prime}_{\rm th}|/\overline{|u^{\prime}_{\rm th}|})$ follow an exponential distribution (Fig. \ref{fig:3}e), compliant with the theory of extreme value statistics \citep{kinnison1983applied,comtet2007level}. Note that we consider the absolute values of $u^{\prime}_{\rm th}$, since their PDFs remain same irrespective of the sign. 

\subsection{Identifying coherent structures}
\label{Res2}
Next, we establish that $u^{\prime}_{\rm th}$ carry the signatures of the outer-layer coherent structures. In wall turbulence, the presence of hairpin structures organize the streamwise velocity field into alternating high- and low-speed streaks \citep{adrian2007hairpin}. This, in turn, induces positive and negative fluctuations in $u^{\prime}$ signals. Through an event synchronization analysis (see Section \ref{ES}), we identify how well these positive and negative patterns are coupled with each other across $y^{+}$. This is achieved through a scale-wise analysis of the Shannon entropies of overlap event lengths, normalized with a RS sequence devoid of any synchronization ($H^{\Delta u_{\rm ref},\Delta u}_{\rm n}(N_p)$, $\Delta u=u^{\prime}(t+\Delta t)-u^{\prime}(t)$). Note that $H^{\Delta u_{\rm ref},\Delta u}_{\rm n}(N_p)$ is bounded between 0 to 1, where 1 (0) indicates no (complete) synchronization. From Fig. \ref{fig:3}f, one observes that the positive and negative events across all $y^{+}$ values are most strongly coupled at scales $\Delta t^{+} \approx 1000$ ($\Delta t^{+}$ is the normalized time lag), representing the outer-layer structures \citep{baars2015wavelet}. This signifies the events in $u^{\prime}$, occurring at heights deep within the inner layer, preserve information about the outer-layer structures.

\begin{figure*}
\centerline{\includegraphics[width=\textwidth]{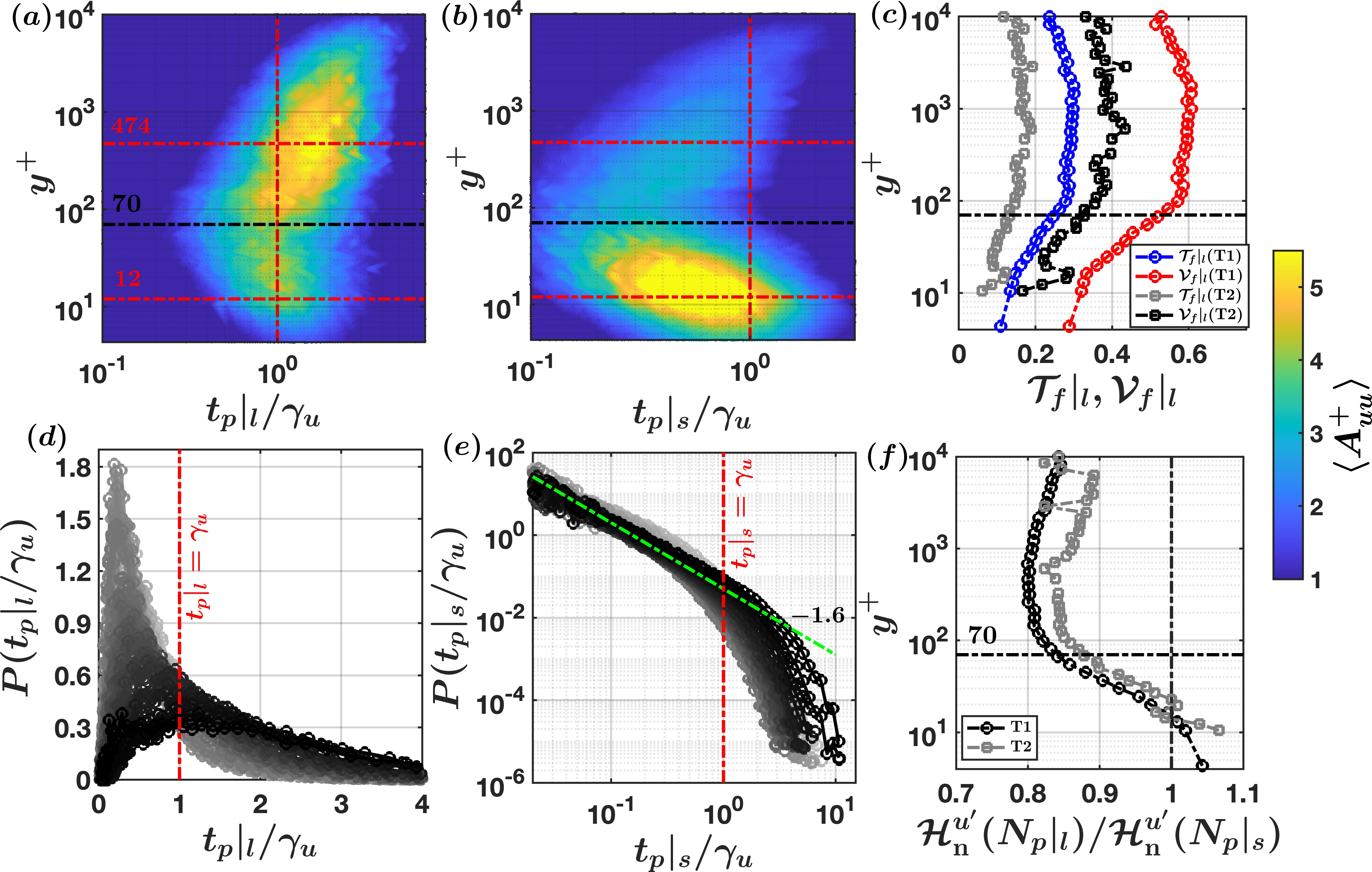}}
\caption{The contours of event amplitudes ($\langle A^{+}_{uu} \rangle$) are plotted separately for the (a) l- and (b) s-type events from T1 dataset. The time scales of these events are denoted as $t_{p}\vert_{l}/\gamma_{u}$ and $t_{p}\vert_s/\gamma_{u}$, where $\gamma_{u}$ is the integral time scale. (c) Fractional contributions of l-type events to the variance ($\mathcal{V}_{f}\vert_{l}$) and occupation time ($\mathcal{T}_{f}\vert_{l}$) are shown. The PDFs of (d) $t_{p}\vert_{l}/\gamma_{u}$ and (e) $t_{p}\vert_s/\gamma_{u}$ are shown from T1 dataset. (f) Shannon entropy ratios of $N_{p}$ corresponding to l- and s-type events are plotted.}  
\label{fig:4}
\end{figure*}

To extract that information, we conditionally sample the events based on whether they contain the samples satisfying $u^{\prime} \geq \alpha^{P}_{\rm th}\sigma_u$ and $u^{\prime} \leq \alpha^{N}_{\rm th}\sigma_u$ (large, or l-type events) or not (small, or s-type events). This concept is graphically illustrated in Appendix \ref{app_B}. The time scales of l- and s-type events are denoted as $t_{p}\vert_{l}/\gamma_{u}$ and $t_{p}\vert_{s}/\gamma_{u}$, respectively. In Figs. \ref{fig:4}a--b, the contributions of l- or s-type events ($\langle A^{+}_{uu} \rangle$, see Eq. (\ref{A_uu}) in Appendix \ref{app_B}) against their time scales to streamwise velocity variance ($\sigma_u^2$) are plotted separately. Quite remarkably, most of the contributions of l-type events to $\sigma_u^2$ come from the heights in and around $y^{+}=474$, where the influence of the outer-layer structures are the strongest \citep{baars2015wavelet}. Conversely, s-type events contribute the most at heights $y^{+}=12$, where the inner-layer structures reside \citep{baars2015wavelet}. Precisely, at heights $y^{+} \geq 70$, the total contributions of l-type events to the velocity variance remain between 40-60\%, although they only occupy $\approx$ 20\% of the time (Fig. \ref{fig:4}c). These contributions compare well with those from VLSMs in wall turbulence \citep{balakumar2007large}. 

Moreover, the PDFs of $t_{p}\vert_{l}/\gamma_{u}$ and $t_{p}\vert_{s}/\gamma_{u}$ appear to be quite different (Figs. \ref{fig:4}d--e). Specifically, $P(t_{p}\vert_{l}/\gamma_{u})$ follows a log-normal distribution (verified with q-q plots), while $P(t_{p}\vert_{s}/\gamma_{u})$ is a power-law of exponent $-1.6$ with an exponential cut off at scales comparable to $\gamma_u$. \citet{sreenivasan2006clustering} demonstrated that the log-normal distribution describes the size distributions of the dissipative structures, while we associate it with the l-type event sizes. Furthermore, by considering the Shannon entropies of event lengths, l-type events are more organized than the s-type ones (since $H^{u^{\prime}}_{\rm n}(N_{p}\vert_{l}) \ll H^{u^{\prime}}_{\rm n}(N_{p}\vert_{s})$) as $y^{+}$ approaches the outer layer (Fig. \ref{fig:4}f). These outcomes confirm that the detected extremes in $u^{\prime}$ carry the signatures of the outer-layer structures and are further utilized to infer about the velocity scales and inner-outer interaction in wall turbulence. Although in Figs. \ref{fig:4}a--b and d--e, T1 dataset is considered, similar findings are obtained for T2 dataset also (see Fig. S1 in Supplementary material).

\subsection{Connections to the turbulent dynamics}
\label{Res3}
\begin{figure*}
\centerline{\includegraphics[width=\textwidth]{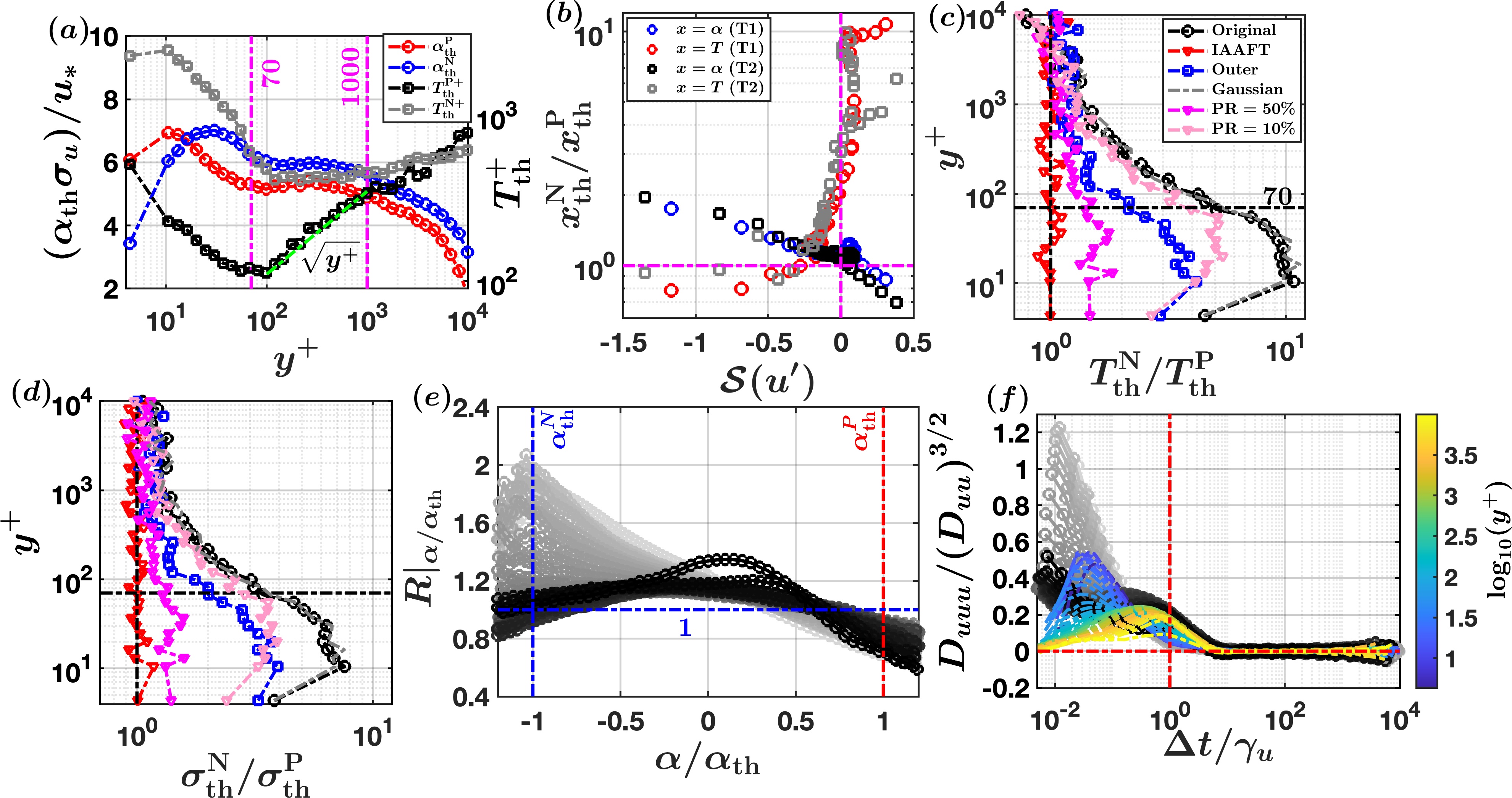}}
\caption{(a) Velocity ($\alpha_{\rm th}\sigma_{u}/u_{*}$) and mean time scales ($T^{+}_{\rm th}$) at the threshold levels are shown for the T1 dataset. (b) Scatter plots between $x^{N}_{\rm th}/x^{P}_{\rm th}$ ($x=\alpha,T$) and $\mathcal{S}(u^{\prime})$ are plotted. For the T1 dataset, ratios of (c) mean time scales ($T^{N}_{\rm th}/T^{P}_{\rm th}$) and (d) their standard deviations ($\sigma^{N}_{\rm th}/\sigma^{P}_{\rm th}$) are compared with Gaussian rank surrogate, phase-randomized surrogate, and with a Fourier-filtered signal for which the outer-layer influences are removed; (e) the mean level-crossing time scales at different $\alpha/\alpha_{\rm th}$-levels are compared with a phase-randomized signal, where $R\vert_{\alpha/\alpha_{\rm th}}$ denotes the ratios between the two; and (f) third-order structure function skewness ($D_{uuu}/{(D_{uu})}^{3/2}$) are compared between the original (gray shaded lines) and a conditionally-shuffled signal (colored lines).}  
\label{fig:5}
\end{figure*}

We construct a velocity scale $(\alpha_{\rm th}\sigma_u)$ for the outer-layer structures and plot their profiles against $y^{+}$ in Fig. \ref{fig:5}a. For $70<y^{+}<10^3$ (i.e., log-layer), this scale attains a near-constant value of $\approx 5u_*$. It remains interesting to see whether this velocity scale, as obtained from the critical $\alpha$ values, can better collapse the turbulence statistics among different experiments. This exercise is, however, out of scope of the present study. On the other hand, to quantify the influence of outer-layer structures on turbulence organization from $u^{\prime}$ time series, we consider the mean time scales at $\alpha=\alpha_{\rm th}$ level ($T_{\rm th}=\overline{t_{p}\vert_{\alpha_{\rm th}}}$). The wall-normalized mean time scales for the positive and negative side are denoted as $T^{P+}_{\rm th}$ and $T^{N+}_{\rm th}$, respectively, with their behaviors being very different. For instance, in the log-layer, $T^{P+}_{\rm th}$ increases as ${(y^{+})}^{1/2}$, while $T^{N+}_{\rm th}$ is nearly constant at $10^{3}$. This increase of $T^{P+}_{\rm th}$ can be explained by considering how the hairpin structures merge progressively to form VLSMs \citep{adrian2007hairpin}, whose characteristic scales ($\approx 10^3$ wall units, \citep{harun2018generation}) match with $T^{N+}_{\rm th}$ values. 

Unlike $\alpha^{N}_{\rm th}/\alpha^{P}_{\rm th}$, the difference between $T^{P}_{\rm th}$ and $T^{N}_{\rm th}$ is anti-correlated to $\mathcal{S}(u')$ (Fig. \ref{fig:5}b). In fact, for both datasets, the largest values of $T^{N}_{\rm th}/T^{P}_{\rm th}$ are obtained when the skewness of $u^{\prime}$ is nearly zero. Instead, we propose that the non-unity values of $T^{N}_{\rm th}/T^{P}_{\rm th}$ are caused by coherent structures and would disappear for a phase-randomized (PR) surrogate, since randomizing the Fourier phases destroys any organizational aspects associated with coherent structures \citep{tobias2008limited}. Clearly, from Fig. \ref{fig:5}c, $T^{N}_{\rm th}/T^{P}_{\rm th}$ approaches 1 for a PR time series (see red dash-dotted line). We use an IAAFT (iteratively adjusted amplitude Fourier transform) model for PR purposes, which preserves the signal PDFs and autocorrelation functions \citep{lancaster2018surrogate}. In fact, if only 10-50\% of the Fourier phases are randomized that itself has a significant effect on $T^{N}_{\rm th}/T^{P}_{\rm th}$ (shown as dash-dotted lines with lighter red shades). Contrarily, if $P(u^{\prime})$ are transformed to Gaussian while maintaining the temporal structure (otherwise known as Gaussian rank surrogate \citep{bogachev2007effect}), then $T^{N}_{\rm th}/T^{P}_{\rm th}$ overlaps with the original (gray dash-dotted line indicates the Gaussian rank surrogate in Fig. \ref{fig:5}c). Hence, the temporal organization of the signal sets the values of $T^{N}_{\rm th}/T^{P}_{\rm th}$. Accordingly, if one removes the outer-layer influences by choosing a Fourier cut-off filter at $\lambda^{+}=7000$ and apply inverse Fourier transform \citep{baars2015wavelet}, it changes $T^{N}_{\rm th}/T^{P}_{\rm th}$ considerably (see blue dash-dotted line in Fig. \ref{fig:5}c). By repeating the analysis on the ratios of the standard deviations or any other higher-order statistics (for instance, skewness and kurtosis) of $t_{p}\vert_{\alpha_{\rm th}}$, the outcome remains the same. However, for illustration purposes, we only show the results in Fig. \ref{fig:5}d corresponding to the standard deviations of $t_{p}\vert_{\alpha_{\rm th}}$ ($\sigma^{N}_{\rm th}/\sigma^{P}_{\rm th}$). Thus, the statistical asymmetry between $t_{p}\vert_{\alpha^{P}_{\rm th}}$ and $t_{p}\vert_{\alpha^{N}_{\rm th}}$ quantifies inner-outer interaction in wall turbulence, as an alternative to the amplitude modulation coefficient proposed by \citet{mathis2009large}. 

Additionally, a PR procedure destroys non-linear dependencies in a signal \citep{lancaster2018surrogate}, which indicates that the non-unity values of $T^{N}_{\rm th}/T^{P}_{\rm th}$ are related to non-linear dynamics. This is at odds with persistence or zero-crossing analysis, where the time scale statistics depend only on the autocorrelation functions accounting for the signals' linear structure \citep{majumdar1999persistence,poggi2009flume}. Therefore, the level-crossing statistics unveil hidden non-linearities in a stochastic signal. To establish this feature more convincingly, in Fig. \ref{fig:5}e, we show how the mean time scales change between the original and PR signal, as $\alpha$ is varied systematically. This is quantified through a timescale ratio, $R\vert_{\alpha/\alpha_{\rm th}}$, as illustrated in Eq. (\ref{ratio}) of Appendix \ref{app_C}. Further this ratio $R\vert_{\alpha/\alpha_{\rm th}}$ deviates from unity, strong non-linear dependencies regulate the timescale statistics. 

Apparently, for heights within the inner layer, non-linear dependencies have the strongest effects on $\overline{t_{p}\vert_{\alpha}}$ at $\alpha_{\rm th}$-level. More importantly, this non-linearity influences $\overline{t_{p}\vert_{\alpha}}$ the most when the threshold is approached from the negative side ($\alpha^{N}_{\rm th}$). Since $\alpha^{N}_{\rm th}$ carry the signatures of the low-speed streaks ($u^{\prime}<0$), this implies that the outer-layer influences on the inner-layer dynamics are governed through a non-linear interaction associated with low-speed streaks. This mechanism was earlier hypothesized by \citet{schoppa2002coherent}, but our results demonstrate it for the first time through an experimental dataset. Interestingly, such non-linear effects on $\overline{t_{p}\vert_{\alpha}}$ become irrelevant when the absolute values of $u^{\prime}$ are considered (see Appendix \ref{app_C}). 

To further investigate the influence of these outer-layer structures on the energy cascading process, we consider a $u^{\prime}$ time series where only the values exceeding $\alpha^{P}_{\rm th}$ and $\alpha^{N}_{\rm th}$ are randomly-shuffled while the others are kept intact. This operation selectively destroys the turbulence organization associated with outer-layer structures. We subsequently calculate the third-order structure function skewness ($D_{uuu}/{(D_{uu})}^{3/2}$), as its non-zero values are related to the turbulence kinetic energy (TKE) cascading from large to small scales \citep{liu2023evolution}. If $D_{uuu}/{(D_{uu})}^{3/2}$ are compared between the original and conditionally-shuffled signal, at scales smaller than $\gamma_u$, $D_{uuu}/{(D_{uu})}^{3/2}$ of the conditionally-shuffled signals decreases significantly with increasing $y^{+}$ (Fig. \ref{fig:5}f). Since within the inner-layer the TKE is also carried by the inner-layer structures, $D_{uuu}/{(D_{uu})}^{3/2}$ values remain slightly larger for the conditionally-shuffled signal. Apart from $D_{uuu}/{(D_{uu})}^{3/2}$ approaching zero, this conditional-shuffling procedure destroys the inertial subrange scaling in second-order structure functions (Fig. S2 in Supplementary material). Therefore, we establish the impact of outer-layer coherent structures on the energy cascade in wall turbulence. It is important to note that these outcomes from Fig. \ref{fig:5} remain unchanged whether T1 or T2 datasets are considered (Fig. S3 in Supplementary material). 

\section{Conclusion}
\label{conclusion}
To summarize, our method of coherent structure detection is entirely data driven and the inferences being obtained from the two datasets match with the existing knowledge of wall-bounded flows. In particular, this detection scheme does not require any external inputs or arbitrary thresholds, thereby making it an attractive choice in experimental turbulence research. This flexibility offers a great advantage in case of atmospheric flows, since coherent structures in such high-$Re$ flows scale with boundary layer height whose measurements are rarely available. Moreover, through level-crossing approach, we provide compelling evidence that the inner-outer interaction in wall turbulence can be quantified by only considering the statistical asymmetry between the peaks and troughs of a turbulent signal. For future research endeavors, it would be interesting to compare this asymmetry parameter among different experiments in wall-bounded turbulence, spanning both internal and external flows. On the interdisciplinary front, the level-crossing framework can be used to detect extremes in other dynamical systems (hydrology, stock markets, etc.), or to generate training datasets for state-of-the-art machine learning models which often fail to predict the extreme occurrences \citep{ding2019modeling}.

\section*{Conflict of Interest Statement}
The authors have no conflicts to disclose.

\section*{Author's contributions}
SC and TB designed and conceptualized this study. SC wrote the manuscript and prepared the figures, while TB provided comments and corrections.

\section*{Supplementary Material}
The Supplementary figures relevant to this article are provided in a separate document.

\section*{Acknowledgements}
SC and TB acknowledge the funding support from the University of California Office of the President (UCOP) grant LFR-20-653572 (UC Lab-Fees); the National Science Foundation (NSF) grants NSF-AGS-PDM-2146520 (CAREER), NSF-OISE-2114740 (AccelNet) and NSF-CPS-2209695 ; the United States Department of Agriculture (USDA) grant 2021-67022-35908 (NIFA); and a cost reimbursable agreement with the USDA Forest Service 20-CR-11242306-072.   

\section*{Availability of data}
The data that support the findings of this study are openly available at \url{https://doi.org/10.26188/5e919e62e0dac}.

\appendix
\appendixpage
\section{Statistical robustness of event entropy curves}
\label{app_A}
\begin{figure*}[h]
\centerline{\includegraphics[width=\textwidth]{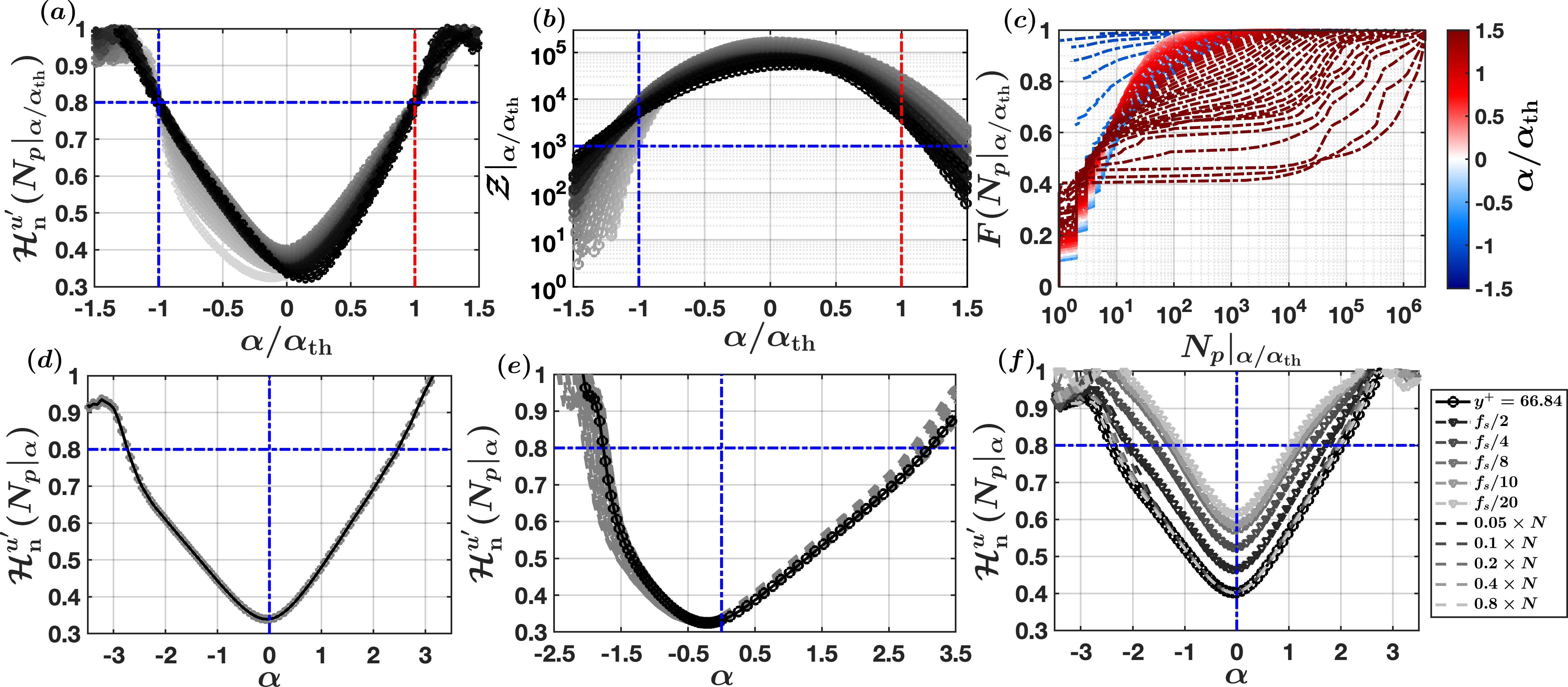}}
\caption{(a) The Shannon entropy curves of event lengths are plotted with respect to the scaled threshold $\alpha/\alpha_{\rm th}$. This scaling ensures that the Shannon entropy curves converge towards the 0.8 value. (b) The number of zero-crossings ($\mathcal{Z}$) are plotted against $\alpha/\alpha_{\rm th}$. The blue horizontal dash-dotted line indicates the number $\mathcal{Z}=10^{3}$. (c) The cumulative distribution functions of the event lengths are shown for different levels of $\alpha/\alpha_{\rm th}$. (d) For $y^{+}=66.84$, the Shannon entropy curve of event lengths (black line) is compared with a randomly-shuffled model of 50 realizations (gray shaded lines). (e) For $y^{+}=4.3$, the Shannon entropy curves are compared between individual ensembles of the measured time series (gray shaded lines) and the averaged one (black line). (f) The impacts of sampling frequencies ($f_s$) and the length of the time series ($N$) on the entropy curve are investigated by systematically varying $f_s$ and $N$ (see the legend).} 
\label{fig:6}
\end{figure*}

We begin by plotting the scaled Shannon entropies of the event lengths (with respect to a RS signal) where $\alpha$ values are normalized with either $\alpha^{P}_{\rm th}$ or $\alpha^{N}_{\rm th}$ (depending on the sign), denoted together as $\alpha_{\rm th}$. Owing to how $\alpha_{\rm th}$ is defined, this normalization ensures that the scaled Shannon entropy curves collapse at 0.8 for all the $y^{+}$ values (Fig. \ref{fig:6}a). However, it raises a question of why we consider 0.8 as our choice instead of 1.

This choice is influenced by the statistical accuracy associated with $H^{u^{\prime}}_{\rm n}$ values. If we consider the mathematical expression of $H^{u^{\prime}}_{\rm n}$, 
\begin{equation}
\mathcal{H}^{u^{\prime}}_{\rm n}(N_{p})=\frac{\sum_{i=1}^{\mathcal{Z}}P(N^{\rm RS}_{p,i})\ln[P(N^{\rm RS}_{p,i})]}{\sum_{i=1}^{\mathcal{Z}}P(N_{p,i})\ln[P(N_{p,i})]},
\label{SN}
\end{equation}
where $\mathcal{Z}$ is the number of times the signal crosses $\alpha$-level, we can clearly see the estimation of $H^{u^{\prime}}_{\rm n}$ is dependent on $\mathcal{Z}$. Our intuition suggests that as $\alpha$ increases, the number of level-crossings would decrease given the rareness in the occurrences of large values in the signal. In Fig. \ref{fig:6}b, we plot the number of level-crossings against $\alpha/\alpha_{\rm th}$ values. As one may note, $\mathcal{Z}$ values decrease beyond 1000 when the $\alpha_{\rm th}$ level is crossed. Since 1000 is a large number to ensure the estimates are statistically robust, we consider the $H^{u^{\prime}}_{\rm n}$ values to be 0.8. This can be further confirmed by plotting the cumulative distribution functions (CDFs) of event lengths. For visualization purposes, we only show the results corresponding to the $u^{\prime}$ signal at $y^{+}=66.84$. Quite clearly, the CDFs display abrupt jumps as $\alpha$ becomes larger than $\alpha_{\rm th}$, due to the lesser number of samples being used to compute their distributions (Fig. \ref{fig:6}c).

It is important to take into account whether the entropy curves when compared with a RS signal change if different realizations of random sequences are used. We test this by generating 50 different realizations of RS sequences and compute the entropy curves for each of such realizations. In Fig. \ref{fig:6}d we show such comparisons using $u^{\prime}$ signal at $y^{+}=66.84$ as the test case. No difference is noted in the results. Moreover, in the figures discussed in the main text, we show only the ensemble-averaged results by combining all the three measurement cycles over which the turbulent time series were collected at each $y^{+}$ value \citep{baars2015wavelet}. In Fig. \ref{fig:6}e, we compare the entropy curves for each ensemble member with the averaged one. We consider the $u^{\prime}$ signal at $y^{+}=4.3$ from the T1 dataset, since at this height the number of ensemble members remains the largest (120). It can be seen that the ensemble-averaged and individual entropy curves almost overlap with no major differences (Fig. \ref{fig:6}e). 

As a last measure, we investigate the influence of the length of the time series ($N$) and sampling frequencies ($f_s$) on the Shannon entropy curves. We artificially change the sampling frequencies by block averaging the $u^{\prime}$ signal values and by doing so we reduce the sampling frequencies as low as 0.05 times the original. Although the entropy curves do change under this operation, their overall shapes remain the same and therefore only appear as a scaled version of the original (Fig. \ref{fig:6}f). This change mainly occurs since by block averaging we alter the standard deviations of the signal and thus the $\alpha$ levels. Potentially it is also possible to increase the sampling frequencies by incorporating an interpolation model, namely piecewise cubic Hermite interpolating polynomial. By utilizing this model, we increase the sampling frequencies two and four times the original, and study its effects on the entropy curves. Similar as before, the curves preserve their shapes and scale according to the $f_s$ values (not shown). On the other hand, if we sub-sample the time series at different lengths compared to the original, $H^{u^{\prime}}_{\rm n}$ remains nearly the same even when sub-sampling reduces the original signal length by 95\% (Fig. \ref{fig:6}f). Hence, we conclude that the estimation of the Shannon entropy curves are statistically robust, placing confidence in the computed $\alpha$ values used later to detect coherent structures.

\section{l- and s-type events}
\label{app_B}
\begin{figure*}[h]
\centerline{\includegraphics[width=\textwidth]{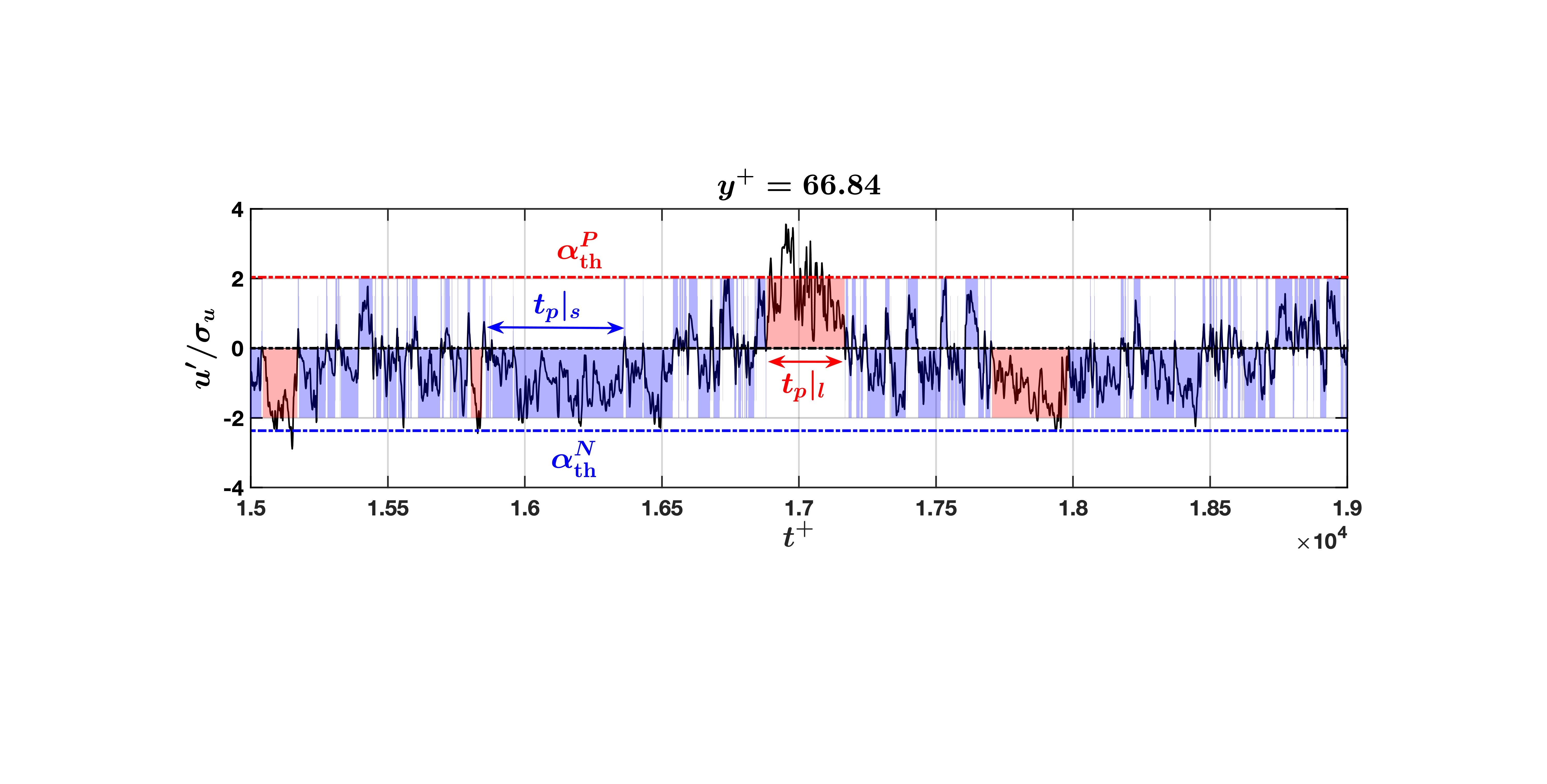}}
\caption{The concept of l- and s-type events are illustrated through a segment of a $u^{\prime}$ time-series at $y^{+}=66.84$ from T1 dataset. The thresholds $\alpha^{P}_{\rm th}$ and $\alpha^{N}_{\rm th}$ identified from the entropy curves (see Fig. \ref{fig:3}b) are shown as horizontal red and blue dash-dotted lines, respectively. The red-colored events (l-type events) contain at least one of these thresholds, whereas the blue-colored ones (s-type events) do not contain any of these. The time scales associated with l- and s-type events are denoted as $t_{p}\vert_l$ and $t_{p}\vert_s$, respectively.} 
\label{fig:7}
\end{figure*}
In this section we illustrate the concepts of l- and s-type events and establish their importance in turbulent dynamics. For this purpose, we use the same segment of the $u^{\prime}$ time series at $y^{+}=66.84$, as done earlier. In Fig. \ref{fig:7}, the three dash-dotted horizontal lines indicate $\alpha=0$ (black), $\alpha^{P}_{\rm th}$ (red), and $\alpha^{N}_{\rm th}$ (blue) levels. The l-type events are defined as those positive or negative blocks where at least one of the $u^{\prime}$ samples satisfy the relation $u^{\prime} \geq \alpha^{P}_{\rm th}\sigma_u$ and $u^{\prime} \leq \alpha^{N}_{\rm th}\sigma_u$. On the other hand, s-type events are those which do not satisfy the above condition. To distinguish the l-type events from s-type ones, we use red-(blue) shaded regions to indicate the l-(s) type events. The time scales associated with l-and s-type events are denoted as $t_{p}\vert_{l}$ and $t_{p}\vert_{s}$ respectively, as shown in Fig. \ref{fig:7}. These time scales are subsequently normalized with $\gamma_u$, which is the integral scale of the $u^{\prime}$ signal. 

Although while demarcating between the l- and s-type events we used $\alpha_{\rm th}$, it is possible to do the same with any $\alpha$ values. For instance, if the $\alpha$ values are chosen to be very small then nearly all the positive and negative events satisfy the condition of the l-type events, and therefore, they become almost indistinguishable from the unconditioned ones (i.e., the original zero-crossing events). Conversely, if the $\alpha$ values are too large then the number of l-type events decrease substantially and overshadowed by the s-type events. Accordingly, it is interesting to consider how the statistics of l- and s-type events change when the $\alpha$ values are varied systematically. We focus on the PDFs of $t_{p}\vert_{l}/\gamma_{u}$ and $t_{p}\vert_{s}/\gamma_{u}$, and the event contributions to the velocity variance. The contributions from a particular event (either l- or s- type) to the velocity variance is defined as,
\begin{equation}
\langle A^{+}_{uu} \rangle=\frac{1}{T \times u_{*}^2}{\int_{t}^{t+(t_{p}\vert_{l,s})} {u^{\prime}}^2(t) \,dt},
\label{A_uu}
\end{equation}
where $T$ is the total signal duration. Note that the contributions are scaled with the friction velocity and further divided by the logarithmic bin-width so the estimations remain nearly independent of the bin choice.

\begin{figure*}[h]
\centerline{\includegraphics[width=\textwidth]{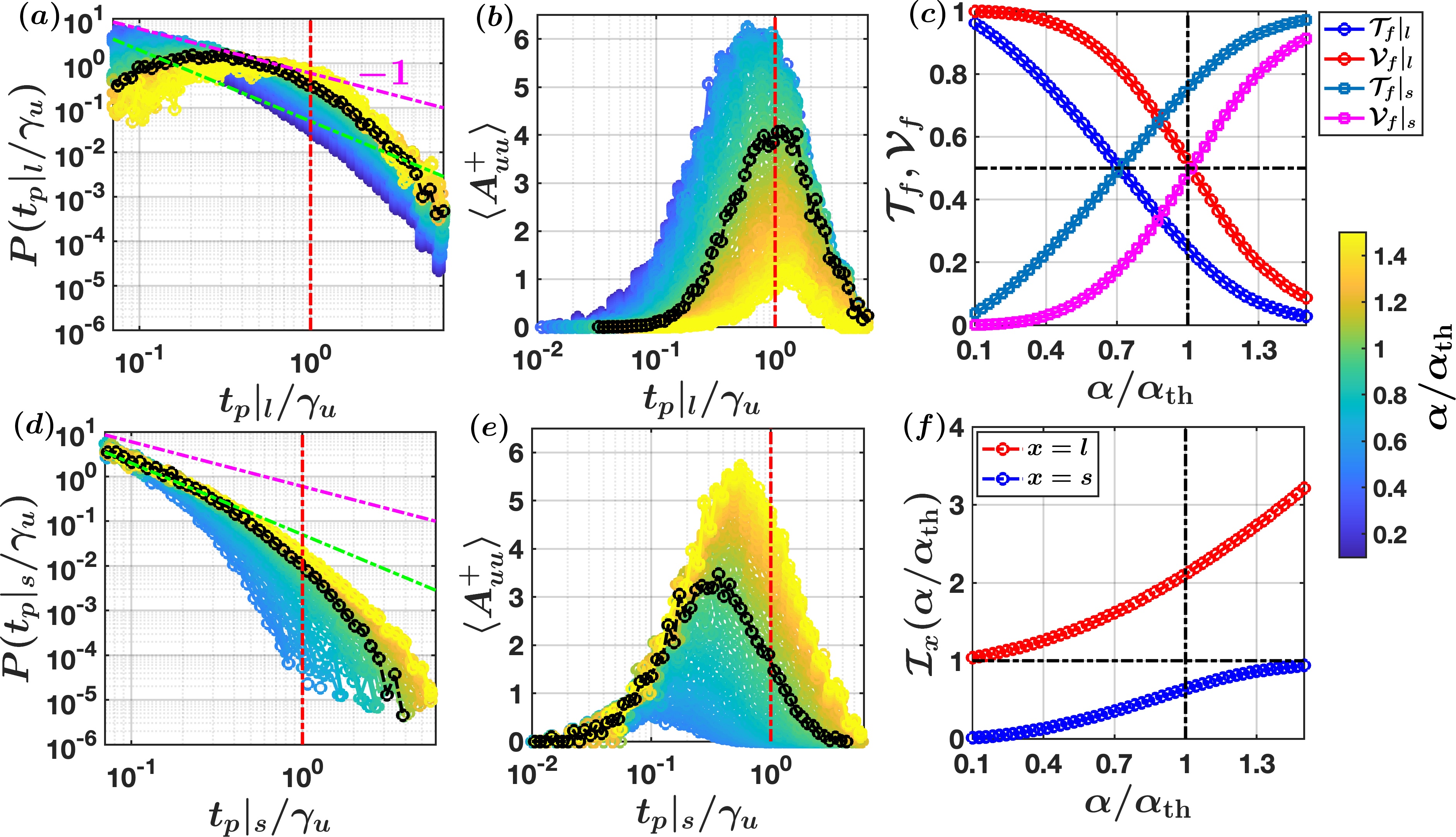}}
\caption{For the l- and s-type events, ((a),(d)) PDFs of time-scales ($P(t_{p}\vert_x/\gamma_{u}$, where $x=\rm l,s$) and the event amplitude ($\langle A^{+}_{uu} \rangle$) curves ((b), (e)) are shown, with systematically changing the $\alpha/\alpha_{\rm th}$ values (see the color bar). The thick black lines denote the curves at $\alpha/\alpha_{\rm th}=1$. The pink dash-dotted lines indicate a power-law of slope $-1$ and the red lines indicate the location where the time scales ($t_{p}\vert_x$) equal to $\gamma_{u}$. (c) Fractional contributions of l-type or s-type events to the variance ($\mathcal{V}_{f}$) and occupation time ($\mathcal{T}_{f}$) are shown for various $\alpha/\alpha_{\rm th}$ values. The vertical and horizontal black lines indicate $\alpha/\alpha_{\rm th}=1$ and $\mathcal{T}_{f},\mathcal{V}_{f}=0.5$. (f) The intermittency coefficient ($\mathcal{I}$) corresponding to l- and s-type events are shown against $\alpha/\alpha_{\rm th}$. } 
\label{fig:8}
\end{figure*}
For the same $u^{\prime}$ signal as used in Fig. \ref{fig:7}, in Figs. \ref{fig:8}a and d, we show how the PDFs of $t_{p}\vert_{l}/\gamma_{u}$ and $t_{p}\vert_{s}/\gamma_{u}$ change as $\alpha$ is varied. Specific to the l-type events, the PDFs at small $\alpha$ values are equivalent to the zero-crossing PDFs of $u^{\prime}$ signal, but as $\alpha$ increases the power-law exponent changes gradually from $-1.6$ to $-1$, with eventually attaining a log-normal distribution. On the other hand, the s-type events approach the zero-crossing PDFs at larger $\alpha$ values, notwithstanding their evolution remains very different from the l-type ones. In particular, the distributions of $t_{p}\vert_{s}/\gamma_{u}$ differ significantly from $t_{p}\vert_{l}/\gamma_{u}$. 

By turning our attention towards event contributions, one can see that with increasing $\alpha$ values the $\langle A^{+}_{uu} \rangle$ curves of l-type events attain their peaks at scales considerably larger than the integral scales (Fig. \ref{fig:8}b). By contrast, the peaks of the $\langle A^{+}_{uu} \rangle$ curves corresponding to s-type events are always smaller than the integral scales (Fig. \ref{fig:8}e). In fact, for small $\alpha$ values, their peaks occur at scales significantly lesser than $\gamma_u$. Therefore, it is plausible that by choosing an appropriate $\alpha$ one might separate the features of small-scale turbulence by conditionally sampling only the s-type events. This is, however, a topic for further research. 

By integrating $\langle A_{uu} \rangle$ curves over all the possible time scales and dividing by the velocity variance, yields fractional contribution to $\sigma_u^2$ ($\mathcal{V}_{f}$) for either of the event types. Similarly, by summing up all the possible time scales and dividing by $T$, yields the occupation time fractions of l- and s-type events ($\mathcal{T}_{f}$). In Fig. \ref{fig:8}c, we show how $\mathcal{T}_{f}$ and $\mathcal{V}_{f}$ vary for the l-type and s-type events against $\alpha/\alpha_{\rm th}$. At $\alpha_{\rm th}$ level, we see that the l-type events nearly contribute 50\% to the velocity variance while occupying 20\% of the time. On the other hand, s-type events occupy 80\% of the time while contributing the same to $\sigma_u^2$. This information can also be studied in terms of an intermittency index ($\mathcal{I}$), defined as a ratio between $\mathcal{V}_{f}$ and $\mathcal{T}_{f}$. 

If $\mathcal{I}$ values are further scaled with the ones obtained from the unconditioned events ($\mathcal{I}_{f}$), then $\mathcal{I}_{f} \to 1$ when $\alpha$ is either too large or small, depending on s- or l-type events respectively. When $\mathcal{I}_{f}$ is plotted against $\alpha/\alpha_{\rm th}$, a clear demarcation is noticed between l- and s-type events in how they approach the unit values (Fig. \ref{fig:8}f). We hypothesize this asymmetrical progression is related to the time-irreversible dynamics of wall-bounded flows \citep{iacobello2023coherent}. 

\section{Sign-indefinite velocity signal}
\label{app_C}
\begin{figure*}[h]
\centerline{\includegraphics[width=\textwidth]{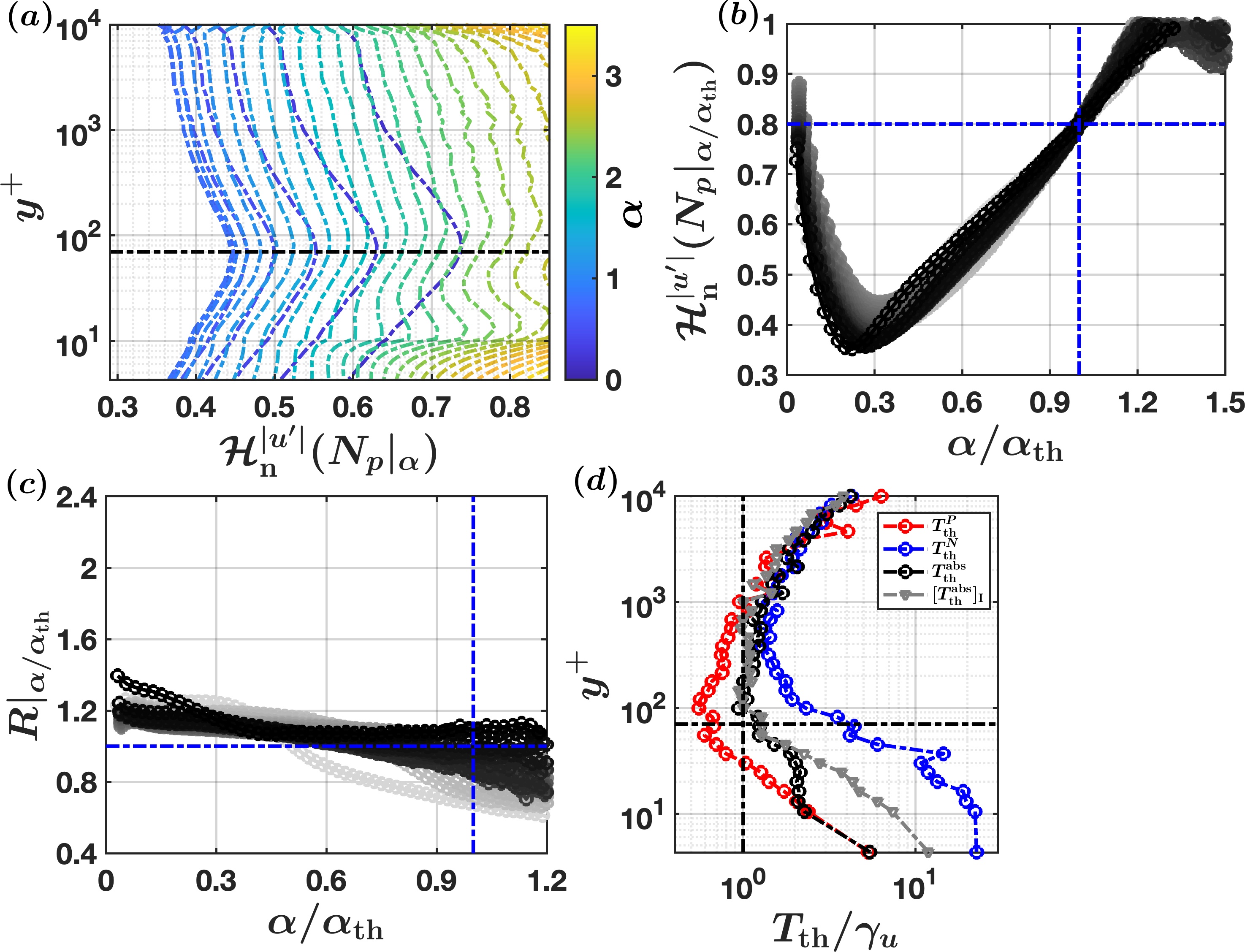}}
\caption{(a) Vertical profiles of the Shannon entropy curves for the absolute values of the $u^{\prime}$ signals ($|u^{\prime}|$) at different $\alpha$-levels are shown for the T1 dataset. (b) $\mathcal{H}^{|u^{\prime}|}_{\rm n}(N_{p}\vert_{\alpha})$ values are plotted for different $y^{+}$ values, and $\alpha_{\rm th}$ is identified where $\mathcal{H}^{|u^{\prime}|}_{\rm n}(N_{p}\vert_{\alpha})=0.8$. (c) The mean level-crossing time scales at different $\alpha/\alpha_{\rm th}$-levels are compared with a phase-randomized signal, where $R\vert_{\alpha/\alpha_{\rm th}}$ denotes the ratios between the two.  (d) The mean time scale computed at $\alpha_{\rm th}$-level ($T_{\rm th}$) are compared between the original and absolute values of the $u^{\prime}$ signals.} 
\label{fig:9}
\end{figure*}
Some earlier studies used thresholds on the time series values to detect coherent structures and suggested that the same could be applied interchangeably on either the original or absolute values of the signal \citep{narasimha2007turbulent}. We, however, show that considering absolute values of the velocity signals instead of the original affects how the events are organized in the temporal space. 

To begin with, we show how the Shannon entropy curves of the event lengths would behave when the $\alpha$ levels are applied on the absolute values of the $u^{\prime}$ signal (Fig. \ref{fig:9}a). Note that it is not possible to set $\alpha=0$ in case of absolute values since no crossings would be obtained in that case. Therefore, the smallest $\alpha$ levels are chosen as slightly larger than 0. By doing so, one observes that up to certain $\alpha$ values the vertical profiles of $H^{|u^{\prime}|}_{\rm n}(N_{p}\vert_{\alpha})$ behave identically as $H^{u^{\prime}}_{\rm n}(N_{p}\vert_{\alpha})$ in Fig. \ref{fig:3}a, when $\alpha$ is approached from the positive side. In fact, similar to $H^{u^{\prime}}_{\rm n}(N_{p}\vert_{\alpha})$, an inflection point in $H^{|u^{\prime}|}_{\rm n}(N_{p}\vert_{\alpha})$ is observed at $y^{+}=70$. 

A note is necessary here regarding the estimation of the critical $\alpha$ value ($\alpha_{\rm th}$) for the $|u^{\prime}|$ signal. The scaled entropy curves of $|u^{\prime}|$ signals form an U shape, and because of that the 0.8 value can be reached either at small or large $\alpha$ levels (Fig. \ref{fig:9}b). At small $\alpha$ levels, the events have large time scales for the absolute signal, since the number of crossings are limited. However, we choose the critical $\alpha$ levels ($\alpha_{\rm th}$) from the larger side, in accordance with the original signal. 

However, the biggest difference between the original and absolute signal occurs when one compares the mean time scales with the phase-randomized (PR) surrogates. This comparison is quantified through a ratio defined as, 
\begin{equation}
R\vert_{\alpha/\alpha_{\rm th}}=\frac{\overline{t_{p}\vert_{\alpha}}}{[{\overline{t_{p}\vert_{\alpha}}}]_{\rm PR}}.
\label{ratio}
\end{equation}
Unlike $u^{\prime}$, for the absolute signals, $R\vert_{\alpha/\alpha_{\rm th}}$ stays almost near to unity for any $\alpha/\alpha_{\rm th}$ values (Fig. \ref{fig:9}c). This indicates, contrary to Fig. \ref{fig:5}e, the effect of non-linear dynamics on the temporal arrangement of the samples exceeding $\alpha_{\rm th}$ disappears by taking the absolute values. We can further confirm this phenomenon by comparing the vertical profiles of $T_{\rm th}/\gamma_{u}$ between $u^{\prime}$ and $|u^{\prime}|$ signals. 

For $|u^{\prime}|$, the mean time scales at $\alpha_{\rm th}$-level remain closer to $T^{P}_{\rm th}$ instead of $T^{N}_{\rm th}$, where $T^{P}_{\rm th}$ and $T^{N}_{\rm th}$ values are obtained from the original $u^{\prime}$ signal (Fig. \ref{fig:9}d). More importantly, $T_{\rm th}$ of the absolute signal is nearly insensitive when the Fourier phases are randomized. Since PR destroys the organization of coherent structures, this indifference suggests that the events detected from the absolute signals may not obey the turbulent flow physics. 

\bibliography{aipsamp}

\providecommand{\noopsort}[1]{}\providecommand{\singleletter}[1]{#1}%
\begin{thebibliography}{41}%
\makeatletter
\providecommand \@ifxundefined [1]{%
 \@ifx{#1\undefined}
}%
\providecommand \@ifnum [1]{%
 \ifnum #1\expandafter \@firstoftwo
 \else \expandafter \@secondoftwo
 \fi
}%
\providecommand \@ifx [1]{%
 \ifx #1\expandafter \@firstoftwo
 \else \expandafter \@secondoftwo
 \fi
}%
\providecommand \natexlab [1]{#1}%
\providecommand \enquote  [1]{``#1''}%
\providecommand \bibnamefont  [1]{#1}%
\providecommand \bibfnamefont [1]{#1}%
\providecommand \citenamefont [1]{#1}%
\providecommand \href@noop [0]{\@secondoftwo}%
\providecommand \href [0]{\begingroup \@sanitize@url \@href}%
\providecommand \@href[1]{\@@startlink{#1}\@@href}%
\providecommand \@@href[1]{\endgroup#1\@@endlink}%
\providecommand \@sanitize@url [0]{\catcode `\\12\catcode `\$12\catcode
  `\&12\catcode `\#12\catcode `\^12\catcode `\_12\catcode `\%12\relax}%
\providecommand \@@startlink[1]{}%
\providecommand \@@endlink[0]{}%
\providecommand \url  [0]{\begingroup\@sanitize@url \@url }%
\providecommand \@url [1]{\endgroup\@href {#1}{\urlprefix }}%
\providecommand \urlprefix  [0]{URL }%
\providecommand \Eprint [0]{\href }%
\providecommand \doibase [0]{http://dx.doi.org/}%
\providecommand \selectlanguage [0]{\@gobble}%
\providecommand \bibinfo  [0]{\@secondoftwo}%
\providecommand \bibfield  [0]{\@secondoftwo}%
\providecommand \translation [1]{[#1]}%
\providecommand \BibitemOpen [0]{}%
\providecommand \bibitemStop [0]{}%
\providecommand \bibitemNoStop [0]{.\EOS\space}%
\providecommand \EOS [0]{\spacefactor3000\relax}%
\providecommand \BibitemShut  [1]{\csname bibitem#1\endcsname}%
\let\auto@bib@innerbib\@empty
\bibitem [{\citenamefont {Kaftori}, \citenamefont {Hetsroni},\ and\
  \citenamefont {Banerjee}(1995)}]{kaftori1995particle}%
  \BibitemOpen
  \bibfield  {author} {\bibinfo {author} {\bibfnamefont {D.}~\bibnamefont
  {Kaftori}}, \bibinfo {author} {\bibfnamefont {G.}~\bibnamefont {Hetsroni}}, \
  and\ \bibinfo {author} {\bibfnamefont {S.}~\bibnamefont {Banerjee}},\
  }\bibfield  {title} {\enquote {\bibinfo {title} {Particle behavior in the
  turbulent boundary layer. {I}. {Motion}, deposition, and entrainment},}\
  }\href@noop {} {\bibfield  {journal} {\bibinfo  {journal} {Phys. Fluids}\
  }\textbf {\bibinfo {volume} {7}},\ \bibinfo {pages} {1095--1106} (\bibinfo
  {year} {1995})}\BibitemShut {NoStop}%
\bibitem [{\citenamefont {Majda}\ and\ \citenamefont
  {Kramer}(1999)}]{majda1999simplified}%
  \BibitemOpen
  \bibfield  {author} {\bibinfo {author} {\bibfnamefont {A.}~\bibnamefont
  {Majda}}\ and\ \bibinfo {author} {\bibfnamefont {P.}~\bibnamefont {Kramer}},\
  }\bibfield  {title} {\enquote {\bibinfo {title} {Simplified models for
  turbulent diffusion: Theory, numerical modelling, and physical phenomena},}\
  }\href@noop {} {\bibfield  {journal} {\bibinfo  {journal} {Phys. Rep.}\
  }\textbf {\bibinfo {volume} {314}},\ \bibinfo {pages} {237--574} (\bibinfo
  {year} {1999})}\BibitemShut {NoStop}%
\bibitem [{\citenamefont {Jim{\'e}nez}(2018)}]{jimenez2018coherent}%
  \BibitemOpen
  \bibfield  {author} {\bibinfo {author} {\bibfnamefont {J.}~\bibnamefont
  {Jim{\'e}nez}},\ }\bibfield  {title} {\enquote {\bibinfo {title} {Coherent
  structures in wall-bounded turbulence},}\ }\href@noop {} {\bibfield
  {journal} {\bibinfo  {journal} {J. Fluid Mech.}\ }\textbf {\bibinfo {volume}
  {842}} (\bibinfo {year} {2018})}\BibitemShut {NoStop}%
\bibitem [{\citenamefont {Chian}\ \emph {et~al.}(2014)\citenamefont {Chian},
  \citenamefont {Rempel}, \citenamefont {Aulanier}, \citenamefont {Schmieder},
  \citenamefont {Shadden}, \citenamefont {Welsch},\ and\ \citenamefont
  {Yeates}}]{chian2014detection}%
  \BibitemOpen
  \bibfield  {author} {\bibinfo {author} {\bibfnamefont {A.~C.-L.}\
  \bibnamefont {Chian}}, \bibinfo {author} {\bibfnamefont {E.~L.}\ \bibnamefont
  {Rempel}}, \bibinfo {author} {\bibfnamefont {G.}~\bibnamefont {Aulanier}},
  \bibinfo {author} {\bibfnamefont {B.}~\bibnamefont {Schmieder}}, \bibinfo
  {author} {\bibfnamefont {S.~C.}\ \bibnamefont {Shadden}}, \bibinfo {author}
  {\bibfnamefont {B.~T.}\ \bibnamefont {Welsch}}, \ and\ \bibinfo {author}
  {\bibfnamefont {A.~R.}\ \bibnamefont {Yeates}},\ }\bibfield  {title}
  {\enquote {\bibinfo {title} {Detection of coherent structures in photospheric
  turbulent flows},}\ }\href@noop {} {\bibfield  {journal} {\bibinfo  {journal}
  {Astrophys. J.}\ }\textbf {\bibinfo {volume} {786}},\ \bibinfo {pages} {51}
  (\bibinfo {year} {2014})}\BibitemShut {NoStop}%
\bibitem [{\citenamefont {Adrian}(2007)}]{adrian2007hairpin}%
  \BibitemOpen
  \bibfield  {author} {\bibinfo {author} {\bibfnamefont {R.~J.}\ \bibnamefont
  {Adrian}},\ }\bibfield  {title} {\enquote {\bibinfo {title} {Hairpin vortex
  organization in wall turbulence},}\ }\href@noop {} {\bibfield  {journal}
  {\bibinfo  {journal} {Phys. Fluids}\ }\textbf {\bibinfo {volume} {19}}
  (\bibinfo {year} {2007})}\BibitemShut {NoStop}%
\bibitem [{\citenamefont {Young}\ \emph {et~al.}(2002)\citenamefont {Young},
  \citenamefont {Kristovich}, \citenamefont {Hjelmfelt},\ and\ \citenamefont
  {Foster}}]{young2002rolls}%
  \BibitemOpen
  \bibfield  {author} {\bibinfo {author} {\bibfnamefont {G.~S.}\ \bibnamefont
  {Young}}, \bibinfo {author} {\bibfnamefont {D.~A.}\ \bibnamefont
  {Kristovich}}, \bibinfo {author} {\bibfnamefont {M.~R.}\ \bibnamefont
  {Hjelmfelt}}, \ and\ \bibinfo {author} {\bibfnamefont {R.~C.}\ \bibnamefont
  {Foster}},\ }\bibfield  {title} {\enquote {\bibinfo {title} {Rolls, streets,
  waves, and more: A review of quasi-two-dimensional structures in the
  atmospheric boundary layer},}\ }\href@noop {} {\bibfield  {journal} {\bibinfo
   {journal} {Bull. Am. Meteorol. Soc.}\ }\textbf {\bibinfo {volume} {83}},\
  \bibinfo {pages} {997--1002} (\bibinfo {year} {2002})}\BibitemShut {NoStop}%
\bibitem [{\citenamefont {Marusic}\ \emph {et~al.}(2021)\citenamefont
  {Marusic}, \citenamefont {Chandran}, \citenamefont {Rouhi}, \citenamefont
  {Fu}, \citenamefont {Wine}, \citenamefont {Holloway}, \citenamefont {Chung},\
  and\ \citenamefont {Smits}}]{marusic2021energy}%
  \BibitemOpen
  \bibfield  {author} {\bibinfo {author} {\bibfnamefont {I.}~\bibnamefont
  {Marusic}}, \bibinfo {author} {\bibfnamefont {D.}~\bibnamefont {Chandran}},
  \bibinfo {author} {\bibfnamefont {A.}~\bibnamefont {Rouhi}}, \bibinfo
  {author} {\bibfnamefont {M.~K.}\ \bibnamefont {Fu}}, \bibinfo {author}
  {\bibfnamefont {D.}~\bibnamefont {Wine}}, \bibinfo {author} {\bibfnamefont
  {B.}~\bibnamefont {Holloway}}, \bibinfo {author} {\bibfnamefont
  {D.}~\bibnamefont {Chung}}, \ and\ \bibinfo {author} {\bibfnamefont {A.~J.}\
  \bibnamefont {Smits}},\ }\bibfield  {title} {\enquote {\bibinfo {title} {An
  energy-efficient pathway to turbulent drag reduction},}\ }\href@noop {}
  {\bibfield  {journal} {\bibinfo  {journal} {Nat. Commun.}\ }\textbf {\bibinfo
  {volume} {12}},\ \bibinfo {pages} {5805} (\bibinfo {year}
  {2021})}\BibitemShut {NoStop}%
\bibitem [{\citenamefont {Salesky}\ and\ \citenamefont
  {Anderson}(2020)}]{salesky2020coherent}%
  \BibitemOpen
  \bibfield  {author} {\bibinfo {author} {\bibfnamefont {S.}~\bibnamefont
  {Salesky}}\ and\ \bibinfo {author} {\bibfnamefont {W.}~\bibnamefont
  {Anderson}},\ }\bibfield  {title} {\enquote {\bibinfo {title} {Coherent
  structures modulate atmospheric surface layer flux-gradient relationships},}\
  }\href@noop {} {\bibfield  {journal} {\bibinfo  {journal} {Phys. Rev. Lett.}\
  }\textbf {\bibinfo {volume} {125}},\ \bibinfo {pages} {124501} (\bibinfo
  {year} {2020})}\BibitemShut {NoStop}%
\bibitem [{\citenamefont {Perry}\ and\ \citenamefont
  {Chong}(1982)}]{perry1982mechanism}%
  \BibitemOpen
  \bibfield  {author} {\bibinfo {author} {\bibfnamefont {A.}~\bibnamefont
  {Perry}}\ and\ \bibinfo {author} {\bibfnamefont {M.}~\bibnamefont {Chong}},\
  }\bibfield  {title} {\enquote {\bibinfo {title} {On the mechanism of wall
  turbulence},}\ }\href@noop {} {\bibfield  {journal} {\bibinfo  {journal} {J.
  Fluid Mech.}\ }\textbf {\bibinfo {volume} {119}},\ \bibinfo {pages}
  {173--217} (\bibinfo {year} {1982})}\BibitemShut {NoStop}%
\bibitem [{\citenamefont {Marusic}\ and\ \citenamefont
  {Monty}(2019)}]{marusic2019attached}%
  \BibitemOpen
  \bibfield  {author} {\bibinfo {author} {\bibfnamefont {I.}~\bibnamefont
  {Marusic}}\ and\ \bibinfo {author} {\bibfnamefont {J.~P.}\ \bibnamefont
  {Monty}},\ }\bibfield  {title} {\enquote {\bibinfo {title} {Attached eddy
  model of wall turbulence},}\ }\href@noop {} {\bibfield  {journal} {\bibinfo
  {journal} {Annu. Rev. Fluid Mech.}\ }\textbf {\bibinfo {volume} {51}},\
  \bibinfo {pages} {49--74} (\bibinfo {year} {2019})}\BibitemShut {NoStop}%
\bibitem [{\citenamefont {Balakumar}\ and\ \citenamefont
  {Adrian}(2007)}]{balakumar2007large}%
  \BibitemOpen
  \bibfield  {author} {\bibinfo {author} {\bibfnamefont {B.}~\bibnamefont
  {Balakumar}}\ and\ \bibinfo {author} {\bibfnamefont {R.}~\bibnamefont
  {Adrian}},\ }\bibfield  {title} {\enquote {\bibinfo {title} {Large-and
  very-large-scale motions in channel and boundary-layer flows},}\ }\href@noop
  {} {\bibfield  {journal} {\bibinfo  {journal} {Philos. Trans. Royal Soc. A}\
  }\textbf {\bibinfo {volume} {365}},\ \bibinfo {pages} {665--681} (\bibinfo
  {year} {2007})}\BibitemShut {NoStop}%
\bibitem [{\citenamefont {Antonia}(1981)}]{antonia1981conditional}%
  \BibitemOpen
  \bibfield  {author} {\bibinfo {author} {\bibfnamefont {R.}~\bibnamefont
  {Antonia}},\ }\bibfield  {title} {\enquote {\bibinfo {title} {Conditional
  sampling in turbulence measurement},}\ }\href@noop {} {\bibfield  {journal}
  {\bibinfo  {journal} {Annu. Rev. Fluid Mech.}\ }\textbf {\bibinfo {volume}
  {13}},\ \bibinfo {pages} {131--156} (\bibinfo {year} {1981})}\BibitemShut
  {NoStop}%
\bibitem [{\citenamefont {Subramanian}\ \emph {et~al.}(1982)\citenamefont
  {Subramanian}, \citenamefont {Rajagopalan}, \citenamefont {Antonia},\ and\
  \citenamefont {Chambers}}]{subramanian1982comparison}%
  \BibitemOpen
  \bibfield  {author} {\bibinfo {author} {\bibfnamefont {C.}~\bibnamefont
  {Subramanian}}, \bibinfo {author} {\bibfnamefont {S.}~\bibnamefont
  {Rajagopalan}}, \bibinfo {author} {\bibfnamefont {R.}~\bibnamefont
  {Antonia}}, \ and\ \bibinfo {author} {\bibfnamefont {A.}~\bibnamefont
  {Chambers}},\ }\bibfield  {title} {\enquote {\bibinfo {title} {Comparison of
  conditional sampling and averaging techniques in a turbulent boundary
  layer},}\ }\href@noop {} {\bibfield  {journal} {\bibinfo  {journal} {J. Fluid
  Mech.}\ }\textbf {\bibinfo {volume} {123}},\ \bibinfo {pages} {335--362}
  (\bibinfo {year} {1982})}\BibitemShut {NoStop}%
\bibitem [{\citenamefont {Alfonsi}(2006)}]{alfonsi2006coherent}%
  \BibitemOpen
  \bibfield  {author} {\bibinfo {author} {\bibfnamefont {G.}~\bibnamefont
  {Alfonsi}},\ }\bibfield  {title} {\enquote {\bibinfo {title} {Coherent
  structures of turbulence: Methods of eduction and results},}\ }\href@noop {}
  {\bibfield  {journal} {\bibinfo  {journal} {Appl. Mech. Rev.}\ }\textbf
  {\bibinfo {volume} {59}} (\bibinfo {year} {2006})}\BibitemShut {NoStop}%
\bibitem [{\citenamefont {Liu}\ and\ \citenamefont
  {Zheng}(2021)}]{liu2021large}%
  \BibitemOpen
  \bibfield  {author} {\bibinfo {author} {\bibfnamefont {H.}~\bibnamefont
  {Liu}}\ and\ \bibinfo {author} {\bibfnamefont {X.}~\bibnamefont {Zheng}},\
  }\bibfield  {title} {\enquote {\bibinfo {title} {Large-scale structures of
  wall-bounded turbulence in single-and two-phase flows: advancing
  understanding of the atmospheric surface layer during sandstorms},}\
  }\href@noop {} {\bibfield  {journal} {\bibinfo  {journal} {Flow}\ }\textbf
  {\bibinfo {volume} {1}} (\bibinfo {year} {2021})}\BibitemShut {NoStop}%
\bibitem [{\citenamefont {Tardu}\ and\ \citenamefont
  {Bauer}(2015)}]{tardu2015level}%
  \BibitemOpen
  \bibfield  {author} {\bibinfo {author} {\bibfnamefont {S.}~\bibnamefont
  {Tardu}}\ and\ \bibinfo {author} {\bibfnamefont {F.}~\bibnamefont {Bauer}},\
  }\bibfield  {title} {\enquote {\bibinfo {title} {Level-crossing statistics
  and production in low {Reynolds} number wall turbulence},}\ }\href@noop {}
  {\bibfield  {journal} {\bibinfo  {journal} {J. Turbul.}\ }\textbf {\bibinfo
  {volume} {16}},\ \bibinfo {pages} {847--871} (\bibinfo {year}
  {2015})}\BibitemShut {NoStop}%
\bibitem [{\citenamefont {Poggi}\ and\ \citenamefont
  {Katul}(2010)}]{poggi2010evaluation}%
  \BibitemOpen
  \bibfield  {author} {\bibinfo {author} {\bibfnamefont {D.}~\bibnamefont
  {Poggi}}\ and\ \bibinfo {author} {\bibfnamefont {G.}~\bibnamefont {Katul}},\
  }\bibfield  {title} {\enquote {\bibinfo {title} {Evaluation of the turbulent
  kinetic energy dissipation rate inside canopies by zero-and level-crossing
  density methods},}\ }\href@noop {} {\bibfield  {journal} {\bibinfo  {journal}
  {Bound.-Layer Meteorol.}\ }\textbf {\bibinfo {volume} {136}},\ \bibinfo
  {pages} {219--233} (\bibinfo {year} {2010})}\BibitemShut {NoStop}%
\bibitem [{\citenamefont {Blake}\ and\ \citenamefont
  {Lindsey}(1973)}]{blake1973level}%
  \BibitemOpen
  \bibfield  {author} {\bibinfo {author} {\bibfnamefont {I.}~\bibnamefont
  {Blake}}\ and\ \bibinfo {author} {\bibfnamefont {W.}~\bibnamefont
  {Lindsey}},\ }\bibfield  {title} {\enquote {\bibinfo {title} {Level-crossing
  problems for random processes},}\ }\href@noop {} {\bibfield  {journal}
  {\bibinfo  {journal} {IEEE Trans. Inf. Theory}\ }\textbf {\bibinfo {volume}
  {19}},\ \bibinfo {pages} {295--315} (\bibinfo {year} {1973})}\BibitemShut
  {NoStop}%
\bibitem [{\citenamefont {Friedrich}\ \emph {et~al.}(2011)\citenamefont
  {Friedrich}, \citenamefont {Peinke}, \citenamefont {Sahimi},\ and\
  \citenamefont {Tabar}}]{friedrich2011approaching}%
  \BibitemOpen
  \bibfield  {author} {\bibinfo {author} {\bibfnamefont {R.}~\bibnamefont
  {Friedrich}}, \bibinfo {author} {\bibfnamefont {J.}~\bibnamefont {Peinke}},
  \bibinfo {author} {\bibfnamefont {M.}~\bibnamefont {Sahimi}}, \ and\ \bibinfo
  {author} {\bibfnamefont {M.~R.~R.}\ \bibnamefont {Tabar}},\ }\bibfield
  {title} {\enquote {\bibinfo {title} {Approaching complexity by stochastic
  methods: From biological systems to turbulence},}\ }\href@noop {} {\bibfield
  {journal} {\bibinfo  {journal} {Phys. Rep.}\ }\textbf {\bibinfo {volume}
  {506}},\ \bibinfo {pages} {87--162} (\bibinfo {year} {2011})}\BibitemShut
  {NoStop}%
\bibitem [{\citenamefont {Majumdar}(1999)}]{majumdar1999persistence}%
  \BibitemOpen
  \bibfield  {author} {\bibinfo {author} {\bibfnamefont {S.~N.}\ \bibnamefont
  {Majumdar}},\ }\bibfield  {title} {\enquote {\bibinfo {title} {Persistence in
  nonequilibrium systems},}\ }\href@noop {} {\bibfield  {journal} {\bibinfo
  {journal} {Curr. Sci.}\ ,\ \bibinfo {pages} {370--375}} (\bibinfo {year}
  {1999})}\BibitemShut {NoStop}%
\bibitem [{\citenamefont {Perlekar}\ \emph {et~al.}(2011)\citenamefont
  {Perlekar}, \citenamefont {Ray}, \citenamefont {Mitra},\ and\ \citenamefont
  {Pandit}}]{perlekar2011persistence}%
  \BibitemOpen
  \bibfield  {author} {\bibinfo {author} {\bibfnamefont {P.}~\bibnamefont
  {Perlekar}}, \bibinfo {author} {\bibfnamefont {S.~S.}\ \bibnamefont {Ray}},
  \bibinfo {author} {\bibfnamefont {D.}~\bibnamefont {Mitra}}, \ and\ \bibinfo
  {author} {\bibfnamefont {R.}~\bibnamefont {Pandit}},\ }\bibfield  {title}
  {\enquote {\bibinfo {title} {Persistence problem in two-dimensional fluid
  turbulence},}\ }\href@noop {} {\bibfield  {journal} {\bibinfo  {journal}
  {Phys. Rev. Lett.}\ }\textbf {\bibinfo {volume} {106}},\ \bibinfo {pages}
  {054501} (\bibinfo {year} {2011})}\BibitemShut {NoStop}%
\bibitem [{\citenamefont {Chowdhuri}, \citenamefont {Kalm{\'a}r-Nagy},\ and\
  \citenamefont {Banerjee}(2020)}]{chowdhuri2020persistence}%
  \BibitemOpen
  \bibfield  {author} {\bibinfo {author} {\bibfnamefont {S.}~\bibnamefont
  {Chowdhuri}}, \bibinfo {author} {\bibfnamefont {T.}~\bibnamefont
  {Kalm{\'a}r-Nagy}}, \ and\ \bibinfo {author} {\bibfnamefont {T.}~\bibnamefont
  {Banerjee}},\ }\bibfield  {title} {\enquote {\bibinfo {title} {Persistence
  analysis of velocity and temperature fluctuations in convective surface layer
  turbulence},}\ }\href@noop {} {\bibfield  {journal} {\bibinfo  {journal}
  {Phys. Fluids}\ }\textbf {\bibinfo {volume} {32}},\ \bibinfo {pages} {076601}
  (\bibinfo {year} {2020})}\BibitemShut {NoStop}%
\bibitem [{\citenamefont {Heisel}\ \emph {et~al.}(2022)\citenamefont {Heisel},
  \citenamefont {de~Silva}, \citenamefont {Katul},\ and\ \citenamefont
  {Chamecki}}]{heisel2022self}%
  \BibitemOpen
  \bibfield  {author} {\bibinfo {author} {\bibfnamefont {M.}~\bibnamefont
  {Heisel}}, \bibinfo {author} {\bibfnamefont {C.~M.}\ \bibnamefont
  {de~Silva}}, \bibinfo {author} {\bibfnamefont {G.~G.}\ \bibnamefont {Katul}},
  \ and\ \bibinfo {author} {\bibfnamefont {M.}~\bibnamefont {Chamecki}},\
  }\bibfield  {title} {\enquote {\bibinfo {title} {Self-similar geometries
  within the inertial subrange of scales in boundary layer turbulence},}\
  }\href@noop {} {\bibfield  {journal} {\bibinfo  {journal} {J. Fluid Mech.}\
  }\textbf {\bibinfo {volume} {942}} (\bibinfo {year} {2022})}\BibitemShut
  {NoStop}%
\bibitem [{\citenamefont {Cava}\ \emph {et~al.}(2012)\citenamefont {Cava},
  \citenamefont {Katul}, \citenamefont {Molini},\ and\ \citenamefont
  {Elefante}}]{cava2012role}%
  \BibitemOpen
  \bibfield  {author} {\bibinfo {author} {\bibfnamefont {D.}~\bibnamefont
  {Cava}}, \bibinfo {author} {\bibfnamefont {G.}~\bibnamefont {Katul}},
  \bibinfo {author} {\bibfnamefont {A.}~\bibnamefont {Molini}}, \ and\ \bibinfo
  {author} {\bibfnamefont {C.}~\bibnamefont {Elefante}},\ }\bibfield  {title}
  {\enquote {\bibinfo {title} {The role of surface characteristics on
  intermittency and zero-crossing properties of atmospheric turbulence},}\
  }\href@noop {} {\bibfield  {journal} {\bibinfo  {journal} {J. Geophys. Res.
  Atmos.}\ }\textbf {\bibinfo {volume} {117}} (\bibinfo {year}
  {2012})}\BibitemShut {NoStop}%
\bibitem [{\citenamefont {Lancaster}\ \emph {et~al.}(2018)\citenamefont
  {Lancaster}, \citenamefont {Iatsenko}, \citenamefont {Pidde}, \citenamefont
  {Ticcinelli},\ and\ \citenamefont {Stefanovska}}]{lancaster2018surrogate}%
  \BibitemOpen
  \bibfield  {author} {\bibinfo {author} {\bibfnamefont {G.}~\bibnamefont
  {Lancaster}}, \bibinfo {author} {\bibfnamefont {D.}~\bibnamefont {Iatsenko}},
  \bibinfo {author} {\bibfnamefont {A.}~\bibnamefont {Pidde}}, \bibinfo
  {author} {\bibfnamefont {V.}~\bibnamefont {Ticcinelli}}, \ and\ \bibinfo
  {author} {\bibfnamefont {A.}~\bibnamefont {Stefanovska}},\ }\bibfield
  {title} {\enquote {\bibinfo {title} {Surrogate data for hypothesis testing of
  physical systems},}\ }\href@noop {} {\bibfield  {journal} {\bibinfo
  {journal} {Phys. Rep.}\ }\textbf {\bibinfo {volume} {748}},\ \bibinfo {pages}
  {1--60} (\bibinfo {year} {2018})}\BibitemShut {NoStop}%
\bibitem [{\citenamefont {Marusic}(2020)}]{MARUSIC2020dataset}%
  \BibitemOpen
  \bibfield  {author} {\bibinfo {author} {\bibfnamefont {I.}~\bibnamefont
  {Marusic}},\ }\href@noop {} {\enquote {\bibinfo {title} {Two-point high
  {Reynolds} number zero-pressure gradient turbulent boundary layer dataset},}\
  }\bibinfo {howpublished} {\url{https://doi.org/10.26188/5e919e62e0dac}}
  (\bibinfo {year} {2020})\BibitemShut {NoStop}%
\bibitem [{\citenamefont {Baars}\ \emph {et~al.}(2015)\citenamefont {Baars},
  \citenamefont {Talluru}, \citenamefont {Hutchins},\ and\ \citenamefont
  {Marusic}}]{baars2015wavelet}%
  \BibitemOpen
  \bibfield  {author} {\bibinfo {author} {\bibfnamefont {W.}~\bibnamefont
  {Baars}}, \bibinfo {author} {\bibfnamefont {K.}~\bibnamefont {Talluru}},
  \bibinfo {author} {\bibfnamefont {N.}~\bibnamefont {Hutchins}}, \ and\
  \bibinfo {author} {\bibfnamefont {I.}~\bibnamefont {Marusic}},\ }\bibfield
  {title} {\enquote {\bibinfo {title} {Wavelet analysis of wall turbulence to
  study large-scale modulation of small scales},}\ }\href@noop {} {\bibfield
  {journal} {\bibinfo  {journal} {Exp. Fluids}\ }\textbf {\bibinfo {volume}
  {56}},\ \bibinfo {pages} {1--15} (\bibinfo {year} {2015})}\BibitemShut
  {NoStop}%
\bibitem [{\citenamefont {Iacobello}\ \emph {et~al.}(2023)\citenamefont
  {Iacobello}, \citenamefont {Chowdhuri}, \citenamefont {Ridolfi},
  \citenamefont {Rondoni},\ and\ \citenamefont
  {Scarsoglio}}]{iacobello2023coherent}%
  \BibitemOpen
  \bibfield  {author} {\bibinfo {author} {\bibfnamefont {G.}~\bibnamefont
  {Iacobello}}, \bibinfo {author} {\bibfnamefont {S.}~\bibnamefont
  {Chowdhuri}}, \bibinfo {author} {\bibfnamefont {L.}~\bibnamefont {Ridolfi}},
  \bibinfo {author} {\bibfnamefont {L.}~\bibnamefont {Rondoni}}, \ and\
  \bibinfo {author} {\bibfnamefont {S.}~\bibnamefont {Scarsoglio}},\ }\bibfield
   {title} {\enquote {\bibinfo {title} {Coherent structures at the origin of
  time irreversibility in wall turbulence},}\ }\href@noop {} {\bibfield
  {journal} {\bibinfo  {journal} {Commun. Phys.}\ }\textbf {\bibinfo {volume}
  {6}},\ \bibinfo {pages} {91} (\bibinfo {year} {2023})}\BibitemShut {NoStop}%
\bibitem [{\citenamefont {Sreenivasan}\ and\ \citenamefont
  {Bershadskii}(2006)}]{sreenivasan2006clustering}%
  \BibitemOpen
  \bibfield  {author} {\bibinfo {author} {\bibfnamefont {K.}~\bibnamefont
  {Sreenivasan}}\ and\ \bibinfo {author} {\bibfnamefont {A.}~\bibnamefont
  {Bershadskii}},\ }\bibfield  {title} {\enquote {\bibinfo {title} {Clustering
  properties in turbulent signals},}\ }\href@noop {} {\bibfield  {journal}
  {\bibinfo  {journal} {J. Stat. Phys.}\ }\textbf {\bibinfo {volume} {125}},\
  \bibinfo {pages} {1141--1153} (\bibinfo {year} {2006})}\BibitemShut {NoStop}%
\bibitem [{\citenamefont {Bae}\ and\ \citenamefont {Lee}(2021)}]{bae2021life}%
  \BibitemOpen
  \bibfield  {author} {\bibinfo {author} {\bibfnamefont {H.~J.}\ \bibnamefont
  {Bae}}\ and\ \bibinfo {author} {\bibfnamefont {M.}~\bibnamefont {Lee}},\
  }\bibfield  {title} {\enquote {\bibinfo {title} {Life cycle of streaks in the
  buffer layer of wall-bounded turbulence},}\ }\href@noop {} {\bibfield
  {journal} {\bibinfo  {journal} {Phys. Rev. Fluids}\ }\textbf {\bibinfo
  {volume} {6}},\ \bibinfo {pages} {064603} (\bibinfo {year}
  {2021})}\BibitemShut {NoStop}%
\bibitem [{\citenamefont {Kinnison}(1983)}]{kinnison1983applied}%
  \BibitemOpen
  \bibfield  {author} {\bibinfo {author} {\bibfnamefont {R.}~\bibnamefont
  {Kinnison}},\ }\href@noop {} {\enquote {\bibinfo {title} {Applied
  extreme-value statistics},}\ }\bibinfo {type} {Tech. Rep.}\ (\bibinfo
  {institution} {Pacific Northwest National Lab.(PNNL), Richland, WA (United
  States)},\ \bibinfo {year} {1983})\BibitemShut {NoStop}%
\bibitem [{\citenamefont {Comtet}, \citenamefont {Leboeuf},\ and\ \citenamefont
  {Majumdar}(2007)}]{comtet2007level}%
  \BibitemOpen
  \bibfield  {author} {\bibinfo {author} {\bibfnamefont {A.}~\bibnamefont
  {Comtet}}, \bibinfo {author} {\bibfnamefont {P.}~\bibnamefont {Leboeuf}}, \
  and\ \bibinfo {author} {\bibfnamefont {S.~N.}\ \bibnamefont {Majumdar}},\
  }\bibfield  {title} {\enquote {\bibinfo {title} {Level density of a {Bose}
  gas and extreme value statistics},}\ }\href@noop {} {\bibfield  {journal}
  {\bibinfo  {journal} {Phys. Rev. Lett.}\ }\textbf {\bibinfo {volume} {98}},\
  \bibinfo {pages} {070404} (\bibinfo {year} {2007})}\BibitemShut {NoStop}%
\bibitem [{\citenamefont {Harun}\ and\ \citenamefont
  {Lotfy}(2018)}]{harun2018generation}%
  \BibitemOpen
  \bibfield  {author} {\bibinfo {author} {\bibfnamefont {Z.}~\bibnamefont
  {Harun}}\ and\ \bibinfo {author} {\bibfnamefont {E.~R.}\ \bibnamefont
  {Lotfy}},\ }\bibfield  {title} {\enquote {\bibinfo {title} {Generation,
  evolution, and characterization of turbulence coherent structures},}\ \
  }(\bibinfo  {publisher} {IntechOpen London, UK},\ \bibinfo {year}
  {2018})\BibitemShut {NoStop}%
\bibitem [{\citenamefont {Tobias}\ and\ \citenamefont
  {Cattaneo}(2008)}]{tobias2008limited}%
  \BibitemOpen
  \bibfield  {author} {\bibinfo {author} {\bibfnamefont {S.~M.}\ \bibnamefont
  {Tobias}}\ and\ \bibinfo {author} {\bibfnamefont {F.}~\bibnamefont
  {Cattaneo}},\ }\bibfield  {title} {\enquote {\bibinfo {title} {Limited role
  of spectra in dynamo theory: Coherent versus random dynamos},}\ }\href@noop
  {} {\bibfield  {journal} {\bibinfo  {journal} {Phys. Rev. Lett.}\ }\textbf
  {\bibinfo {volume} {101}},\ \bibinfo {pages} {125003} (\bibinfo {year}
  {2008})}\BibitemShut {NoStop}%
\bibitem [{\citenamefont {Bogachev}, \citenamefont {Eichner},\ and\
  \citenamefont {Bunde}(2007)}]{bogachev2007effect}%
  \BibitemOpen
  \bibfield  {author} {\bibinfo {author} {\bibfnamefont {M.~I.}\ \bibnamefont
  {Bogachev}}, \bibinfo {author} {\bibfnamefont {J.~F.}\ \bibnamefont
  {Eichner}}, \ and\ \bibinfo {author} {\bibfnamefont {A.}~\bibnamefont
  {Bunde}},\ }\bibfield  {title} {\enquote {\bibinfo {title} {Effect of
  nonlinear correlations on the statistics of return intervals in multifractal
  data sets},}\ }\href@noop {} {\bibfield  {journal} {\bibinfo  {journal}
  {Phys. Rev. Lett.}\ }\textbf {\bibinfo {volume} {99}},\ \bibinfo {pages}
  {240601} (\bibinfo {year} {2007})}\BibitemShut {NoStop}%
\bibitem [{\citenamefont {Mathis}, \citenamefont {Hutchins},\ and\
  \citenamefont {Marusic}(2009)}]{mathis2009large}%
  \BibitemOpen
  \bibfield  {author} {\bibinfo {author} {\bibfnamefont {R.}~\bibnamefont
  {Mathis}}, \bibinfo {author} {\bibfnamefont {N.}~\bibnamefont {Hutchins}}, \
  and\ \bibinfo {author} {\bibfnamefont {I.}~\bibnamefont {Marusic}},\
  }\bibfield  {title} {\enquote {\bibinfo {title} {Large-scale amplitude
  modulation of the small-scale structures in turbulent boundary layers},}\
  }\href@noop {} {\bibfield  {journal} {\bibinfo  {journal} {J. Fluid Mech.}\
  }\textbf {\bibinfo {volume} {628}},\ \bibinfo {pages} {311--337} (\bibinfo
  {year} {2009})}\BibitemShut {NoStop}%
\bibitem [{\citenamefont {Poggi}\ and\ \citenamefont
  {Katul}(2009)}]{poggi2009flume}%
  \BibitemOpen
  \bibfield  {author} {\bibinfo {author} {\bibfnamefont {D.}~\bibnamefont
  {Poggi}}\ and\ \bibinfo {author} {\bibfnamefont {G.}~\bibnamefont {Katul}},\
  }\bibfield  {title} {\enquote {\bibinfo {title} {Flume experiments on
  intermittency and zero-crossing properties of canopy turbulence},}\
  }\href@noop {} {\bibfield  {journal} {\bibinfo  {journal} {Phys. Fluids}\
  }\textbf {\bibinfo {volume} {21}} (\bibinfo {year} {2009})}\BibitemShut
  {NoStop}%
\bibitem [{\citenamefont {Schoppa}\ and\ \citenamefont
  {Hussain}(2002)}]{schoppa2002coherent}%
  \BibitemOpen
  \bibfield  {author} {\bibinfo {author} {\bibfnamefont {W.}~\bibnamefont
  {Schoppa}}\ and\ \bibinfo {author} {\bibfnamefont {F.}~\bibnamefont
  {Hussain}},\ }\bibfield  {title} {\enquote {\bibinfo {title} {Coherent
  structure generation in near-wall turbulence},}\ }\href@noop {} {\bibfield
  {journal} {\bibinfo  {journal} {J. Fluid Mech.}\ }\textbf {\bibinfo {volume}
  {453}},\ \bibinfo {pages} {57--108} (\bibinfo {year} {2002})}\BibitemShut
  {NoStop}%
\bibitem [{\citenamefont {Liu}\ \emph {et~al.}(2023)\citenamefont {Liu},
  \citenamefont {Wu}, \citenamefont {Lv}, \citenamefont {Yang},\ and\
  \citenamefont {Zhou}}]{liu2023evolution}%
  \BibitemOpen
  \bibfield  {author} {\bibinfo {author} {\bibfnamefont {F.}~\bibnamefont
  {Liu}}, \bibinfo {author} {\bibfnamefont {Z.}~\bibnamefont {Wu}}, \bibinfo
  {author} {\bibfnamefont {P.}~\bibnamefont {Lv}}, \bibinfo {author}
  {\bibfnamefont {W.}~\bibnamefont {Yang}}, \ and\ \bibinfo {author}
  {\bibfnamefont {Y.}~\bibnamefont {Zhou}},\ }\bibfield  {title} {\enquote
  {\bibinfo {title} {Evolution of the velocity gradient invariants in
  homogeneous isotropic turbulence with an inverse energy cascade},}\
  }\href@noop {} {\bibfield  {journal} {\bibinfo  {journal} {Phys. Fluids}\
  }\textbf {\bibinfo {volume} {35}} (\bibinfo {year} {2023})}\BibitemShut
  {NoStop}%
\bibitem [{\citenamefont {Ding}\ \emph {et~al.}(2019)\citenamefont {Ding},
  \citenamefont {Zhang}, \citenamefont {Pan}, \citenamefont {Yang},\ and\
  \citenamefont {He}}]{ding2019modeling}%
  \BibitemOpen
  \bibfield  {author} {\bibinfo {author} {\bibfnamefont {D.}~\bibnamefont
  {Ding}}, \bibinfo {author} {\bibfnamefont {M.}~\bibnamefont {Zhang}},
  \bibinfo {author} {\bibfnamefont {X.}~\bibnamefont {Pan}}, \bibinfo {author}
  {\bibfnamefont {M.}~\bibnamefont {Yang}}, \ and\ \bibinfo {author}
  {\bibfnamefont {X.}~\bibnamefont {He}},\ }\bibfield  {title} {\enquote
  {\bibinfo {title} {Modeling extreme events in time series prediction},}\ }in\
  \href@noop {} {\emph {\bibinfo {booktitle} {Proceedings of the 25th ACM
  SIGKDD International Conference on Knowledge Discovery \& Data Mining}}}\
  (\bibinfo {year} {2019})\ pp.\ \bibinfo {pages} {1114--1122}\BibitemShut
  {NoStop}%
\bibitem [{\citenamefont {Narasimha}\ \emph {et~al.}(2007)\citenamefont
  {Narasimha}, \citenamefont {Kumar}, \citenamefont {Prabhu},\ and\
  \citenamefont {Kailas}}]{narasimha2007turbulent}%
  \BibitemOpen
  \bibfield  {author} {\bibinfo {author} {\bibfnamefont {R.}~\bibnamefont
  {Narasimha}}, \bibinfo {author} {\bibfnamefont {S.}~\bibnamefont {Kumar}},
  \bibinfo {author} {\bibfnamefont {A.}~\bibnamefont {Prabhu}}, \ and\ \bibinfo
  {author} {\bibfnamefont {S.}~\bibnamefont {Kailas}},\ }\bibfield  {title}
  {\enquote {\bibinfo {title} {Turbulent flux events in a nearly neutral
  atmospheric boundary layer},}\ }\href@noop {} {\bibfield  {journal} {\bibinfo
   {journal} {Phil. Trans. R. Soc. A}\ }\textbf {\bibinfo {volume} {365}},\
  \bibinfo {pages} {841--858} (\bibinfo {year} {2007})}\BibitemShut {NoStop}%
\end{thebibliography}%

\end{document}